\documentclass[12pt,preprint]{aastex}
\usepackage{lscape}
\usepackage{rotating}
\def\rel{{\rm rel}}
\def\min{{\rm min}}
\def\max{{\rm max}}
\def\thr{{\rm thr}}
\def\pl{{\rm pl}}
\def\ev{{\rm ev}}
\def\e{{\rm E}}
\def\hbn{{\hfil\break\noindent}}
\begin{document}
\title{Frequency of Solar-Like Systems and of
Ice and Gas Giants Beyond the Snow Line
from High-Magnification Microlensing Events in 2005-2008}

\author{
A.~Gould\altaffilmark{1,2}, 
Subo~Dong\altaffilmark{1,3,4},
B.S.~Gaudi\altaffilmark{1,2},
A. Udalski\altaffilmark{5,6},
I.A.~Bond\altaffilmark{7,8},
J.~Greenhill\altaffilmark{9,10},
R.A.~Street\altaffilmark{11,12,13},   
M. Dominik\altaffilmark{9,11,14,15,16},
T.~Sumi\altaffilmark{7,17},
M.K. Szyma{\'n}ski\altaffilmark{5,6}, 
C.~Han\altaffilmark{1,18},\\
and\\
W.~Allen\altaffilmark{19},
G.~Bolt\altaffilmark{20},
M.~Bos\altaffilmark{21},
G.W.~Christie\altaffilmark{22},
D.L.~DePoy\altaffilmark{23},
J.~Drummond\altaffilmark{24},
J.D.~Eastman\altaffilmark{2},
A.~Gal-Yam\altaffilmark{25},
D.~Higgins\altaffilmark{26},
J.~Janczak\altaffilmark{2},
S.~Kaspi\altaffilmark{27,28},
S.~Koz{\l}owski\altaffilmark{2},
C.-U.~Lee\altaffilmark{29},
F.~Mallia\altaffilmark{30},
A.~Maury\altaffilmark{30},
D.~Maoz\altaffilmark{27},
J.~McCormick\altaffilmark{31},
L.A.G.~Monard\altaffilmark{32},
D.~Moorhouse\altaffilmark{33},
N.~Morgan\altaffilmark{2},
T.~Natusch\altaffilmark{34},
E.O.~Ofek\altaffilmark{35,36},
B.-G.~Park\altaffilmark{29},
R.W.~Pogge\altaffilmark{2},
D.~Polishook\altaffilmark{27},
R.~Santallo\altaffilmark{37},
A.~Shporer\altaffilmark{27},
O.~Spector\altaffilmark{27},
G.~Thornley\altaffilmark{33},
J.C.~Yee\altaffilmark{2}\\
({The $\mu$FUN Collaboration}),\\
and\\
M. Kubiak\altaffilmark{6},
G. Pietrzy{\'n}ski\altaffilmark{6,38}, 
I. Soszy{\'n}ski\altaffilmark{6},
O. Szewczyk\altaffilmark{38}, 
{\L}. Wyrzykowski\altaffilmark{39}, 
K. Ulaczyk\altaffilmark{6},
R. Poleski\altaffilmark{6}\\
({The OGLE Collaboration}),\\
and\\
F.~Abe\altaffilmark{17},
D.P.~Bennett\altaffilmark{40,9},
C.S.~Botzler\altaffilmark{41},
D.~Douchin\altaffilmark{41},
M.~Freeman\altaffilmark{41},
A.~Fukui\altaffilmark{17},
K.~Furusawa\altaffilmark{17},
J.B.~Hearnshaw\altaffilmark{42},
S.~Hosaka\altaffilmark{17},
Y.~Itow\altaffilmark{17},
K.~Kamiya\altaffilmark{17},
P.M.~Kilmartin\altaffilmark{43},
A.~Korpela\altaffilmark{44},
W.~Lin\altaffilmark{8},
C.H.~Ling\altaffilmark{8},
S.~Makita\altaffilmark{17},
K.~Masuda\altaffilmark{17},
Y.~Matsubara\altaffilmark{17},
N.~Miyake\altaffilmark{17},
Y.~Muraki\altaffilmark{45},
M.~Nagaya\altaffilmark{17},
K.~Nishimoto\altaffilmark{17}
K.~Ohnishi\altaffilmark{46},
T.~Okumura\altaffilmark{17},
Y.C.~Perrott\altaffilmark{41},
L.~Philpott\altaffilmark{41},
N.~Rattenbury\altaffilmark{41,11},
To.~Saito\altaffilmark{47},
T.~Sako\altaffilmark{17},
D.J.~Sullivan\altaffilmark{44},
W.L.~Sweatman\altaffilmark{8},
P.J.~Tristram\altaffilmark{43},
E.~von Seggern\altaffilmark{41},
P.C.M.~Yock\altaffilmark{41}\\
({The MOA Collaboration}),\\
and\\
M.~Albrow\altaffilmark{42},
V.~Batista\altaffilmark{48},
J.P.~Beaulieu\altaffilmark{48},
S.~Brillant\altaffilmark{49},   
J.~Caldwell\altaffilmark{50},
J.J.~Calitz\altaffilmark{51},
A.~Cassan\altaffilmark{48},
A.~Cole\altaffilmark{10},
K.~Cook\altaffilmark{52},
C.~Coutures\altaffilmark{53},
S.~Dieters\altaffilmark{48},  
D.~Dominis Prester\altaffilmark{54},
J.~Donatowicz\altaffilmark{55},
P.~Fouqu\'{e}\altaffilmark{56},
K.~Hill\altaffilmark{10},
M.~Hoffman\altaffilmark{51},
F.~Jablonski\altaffilmark{57},
S.R.~Kane\altaffilmark{58},
N.~Kains\altaffilmark{15,11},
D.~Kubas\altaffilmark{49},
J.-B.~Marquette\altaffilmark{48},   
R.~Martin\altaffilmark{59},
E.~Martioli\altaffilmark{57},
P.~Meintjes\altaffilmark{51},
J.~Menzies\altaffilmark{60},
E.~Pedretti\altaffilmark{15},  
K.~Pollard\altaffilmark{42},
K.C.~Sahu\altaffilmark{61},
C.~Vinter\altaffilmark{62},
J.~Wambsganss\altaffilmark{63,14},
R.~Watson\altaffilmark{10},
A.~Williams\altaffilmark{59},
M. Zub\altaffilmark{63,64}\\
({The PLANET Collaboration}),\\
and\\
A.~Allan\altaffilmark{65},   
M.F.~Bode\altaffilmark{66},   
D.M.~Bramich\altaffilmark{67,9},   
M.J.~Burgdorf\altaffilmark{68,69,14},   
N.~Clay\altaffilmark{66},   
S.~Fraser\altaffilmark{66},   
E.~Hawkins\altaffilmark{12},
K.~Horne\altaffilmark{15,9},   
E.~Kerins\altaffilmark{70},
T.A.~Lister\altaffilmark{12},   
C.~Mottram\altaffilmark{66},   
E.S.~Saunders\altaffilmark{12,65},   
C.~Snodgrass\altaffilmark{49,71,14},   
I.A.~Steele\altaffilmark{66},   
Y.~Tsapras\altaffilmark{12,72,9}\\
({The RoboNet Collaboration}),\\
and\\
U. G. J{\o}rgensen\altaffilmark{62,73,9}, 
T. Anguita\altaffilmark{63,74},
V. Bozza\altaffilmark{75,76}, 
S. Calchi Novati\altaffilmark{75,76},
K. Harps{\o}e\altaffilmark{62},
T. C. Hinse\altaffilmark{62,77}, 
M. Hundertmark\altaffilmark{78},
P. Kj{\ae}rgaard\altaffilmark{62},
C. Liebig\altaffilmark{15,63}, 
L. Mancini\altaffilmark{75,76},
G. Masi\altaffilmark{79}, 
M. Mathiasen\altaffilmark{62},
S. Rahvar\altaffilmark{80}, 
D. Ricci\altaffilmark{81},
G. Scarpetta\altaffilmark{75,76}, 
J. Southworth\altaffilmark{82},
J. Surdej\altaffilmark{81}, 
C. C. Th\"{o}ne\altaffilmark{62,83}\\
({The MiNDSTEp Consortium})\\
}

\altaffiltext{1}{Microlensing Follow Up Network ($\mu$FUN)}
\altaffiltext{2}{Department of Astronomy, Ohio State University,
140 W.\ 18th Ave., Columbus, OH 43210, USA; 
gould,gaudi,jdeast,jyee,pogge,simkoz@astronomy.ohio-state.edu, nick.morgan@alum.mit.edu}
\altaffiltext{3}{Institute for Advanced Study, Einstein Drive, Princeton, NJ 08540, USA; dong@ias.edu}
\altaffiltext{4}{Sagan Fellow}
\altaffiltext{5}{Optical Gravitational Lens Experiment (OGLE)}
\altaffiltext{6}{Warsaw University Observatory, Al. Ujazdowskie 4, 00-478
Warszawa, Poland; e-mail: udalski, msz, mk, pietrzyn,
soszynsk, kulaczyk, rpoleski@astrouw.edu.pl}
\altaffiltext{7}{Microlensing Observations in Astrophysics (MOA)}
\altaffiltext{8}
{Institute of Information and Mathematical Sciences, Massey University,
Private Bag 102-904, North Shore Mail Centre, Auckland, New Zealand;
i.a.bond,l.skuljan,w.lin,c.h.ling,w.sweatman@massey.ac.nz}
\altaffiltext{9}{Probing Lensing Anomalies NETwork (PLANET)} 
\altaffiltext{10}{University
of Tasmania, School of Mathematics and Physics, Private Bag 37, GPO
Hobart, Tas 7001, Australia; John.Greenhill,Andrew.Cole@utas.edu.au	}
\altaffiltext{11}{RoboNet} 
\altaffiltext{12}{Las Cumbres Observatory Global Telescope network, 6740 Cortona Drive, suite 102, Goleta, CA 93117, USA}
\altaffiltext{13}{Dept. of Physics, Broida Hall, University of California, Santa Barbara CA 93106-9530, USA}
\altaffiltext{14}{Microlensing Network for the
Detection of Small Terrestrial Exoplanets (MiNDSTEp)}
\altaffiltext{15}{SUPA School of Physics and Astronomy, Univ. of St Andrews, Scotland KY16 9SS, United Kingdom; md35,nk87,ep41,kdh1@st-andrews.ac.uk}
\altaffiltext{16}{Royal Society University Research Fellow}

\altaffiltext{17}{Solar-Terrestrial Environment Laboratory, Nagoya University, Nagoya, 464-8601, Japan; sumi,abe,afukui,furusawa,itow,kkamiya,kmasuda,ymatsu,nmiyake,mnagaya,okumurat,sako@stelab.nagoya-u.ac.jp}
\altaffiltext{18}
{Department of Physics, Institute for Basic Science Research,
Chungbuk National University, Chongju 361-763, Korea;
cheongho@astroph.chungbuk.ac.kr}
\altaffiltext{19}
{Vintage Lane Observatory, Blenheim, New Zealand; whallen@xtra.co.nz}
\altaffiltext{20}{Perth, Australia; gbolt@iinet.net.au}

\altaffiltext{21}
{Molehill Astronomical Observatory, Auckland, New Zealand; molehill@ihug.co.nz}
\altaffiltext{22}{Auckland Observatory, Auckland, New Zealand; gwchristie@christie.org.nz}
\altaffiltext{23}
{Department of Physics and Astronomy, Texas A\&M University, College Station, TX, USA; depoy@physics.tamu.edu}
\altaffiltext{24}{Possum Observatory, Patutahi, New Zealand; john\_drummond@xtra.co.nz}
\altaffiltext{25}{Department of Particle Physics and Astrophysics,
Weizmann Institute of Science,
76100 Rehovot,
Israel; avishay.gal-yam@weizmann.ac.il}
\altaffiltext{26}
{Hunters Hill Observatory, Canberra, Australia; dhi67540@bigpond.net.au}
\altaffiltext{27}{School of Physics and Astronomy and Wise Observatory, Tel-Aviv University, Tel-Aviv 69978, Israel; shai,dani,david,shporer,odedspec@wise.tau.ac.il}
\altaffiltext{28}{Department of Physics, Technion, Haifa 32000, Israel}
\altaffiltext{29}
{Korea Astronomy and
Space Science Institute, Daejon 305-348, Korea; leecu,bgpark@kasi.re.kr}
\altaffiltext{30}
{Campo Catino Austral Observatory, San Pedro de Atacama, Chile; francomallia@campocatinobservatory.org,alain@spaceobs.com}
\altaffiltext{31}
{Farm Cove Observatory, Centre for Backyard Astrophysics,
Pakuranga, Auckland, New Zealand; farmcoveobs@xtra.co.nz}
\altaffiltext{32}
{Bronberg Observatory, Centre for Backyard Astrophysics, Pretoria, South
Africa; lagmonar@nmisa.org}
\altaffiltext{33}{Kumeu Observatory, Kumeu, New Zealand; acrux@orcon.net.nz,guy.thornley@gmail.com}
\altaffiltext{34}
{AUT University, Auckland, New Zealand; tim.natusch@aut.ac.nz}
\altaffiltext{35}{Palomar Observatory, California, USA; eran@astro.caltech.edu}
\altaffiltext{36}{Einstein Fellow}
\altaffiltext{37}
{Southern Stars Observatory, Faaa, Tahiti, French Polynesia; obs930@southernstars-observatory.org}
\altaffiltext{38}{Universidad de Concepci{\'o}n, Departamento de Fisica, Casilla 160--C, Concepci{\'o}n, Chile; e-mail: szewczyk@astro-udec.cl}
\altaffiltext{39}{Institute of Astronomy, University of Cambridge, Madingley Road,
CB3 0HA Cambridge, United Kingdom; e-mail: wyrzykow@ast.cam.ac.uk}
\altaffiltext{40}{University of Notre Dame, Department of Physics, 225 Nieuwland Science Hall, Notre Dame, IN 46556-5670 USA; bennett@nd.edu} 

\altaffiltext{41}
{Department of Physics, University of Auckland, Private Bag 92019, Auckland, New Zealand; c.botzler,p.yock@auckland.ac.nz,yper006@aucklanduni.ac.nz}
\altaffiltext{42}
{University of Canterbury, Department of Physics and Astronomy, Private Bag 4800, Christchurch 8020, New Zealand}
\altaffiltext{43}
{Mt.\ John Observatory, P.O. Box 56, Lake Tekapo 8780, New Zealand}
\altaffiltext{44}
{School of Chemical and Physical Sciences, Victoria University, Wellington, New Zealand; a.korpela@niwa.co.nz,denis.sullivan@vuw.ac.nz}
\altaffiltext{45}
{Department of Physics, Konan University, Nishiokamoto 8-9-1, Kobe 658-8501, Japan}
\altaffiltext{46}
{Nagano National College of Technology, Nagano 381-8550, Japan}
\altaffiltext{47}
{Tokyo Metropolitan College of Industrial Technology, Tokyo 116-8523, Japan}
\altaffiltext{48}{Institut d'Astrophysique de Paris, Universit{\'e} Pierre et Marie Curie, CNRS UMR7095, 98bis Boulevard Arago, 75014 Paris, France; beaulieu,marquett@iap.fr}
\altaffiltext{49}{European Southern Observatory, Casilla 19001, Santiago 19, Chile; sbrillan,dkubas@eso.org}
\altaffiltext{50}{McDonald Observatory, 16120 St Hwy Spur 78 \#2, Fort Davis, TX 79734 USA; caldwell@astro.as.utexas.edu}
\altaffiltext{51}{University of the Free State, Faculty of Natural and Agricultural Sciences, Department of Physics, PO Box 339, Bloemfontein 9300, South Africa; HoffmaMJ.SCI@mail.uovs.ac.za}
\altaffiltext{52}{Lawrence Livermore National Laboratory, Institute of Geophysics and Planetary Physics, P.O. Box 808, Livermore, CA 94551-0808 USA; kcook@llnl.gov}
\altaffiltext{53}{CEA/Saclay, 91191 Gif-sur-Yvette cedex, France; coutures@iap.fr}
\altaffiltext{54}{Department of Physics, University of Rijeka, Omladinska 14, 51000 Rijeka, Croatia}
\altaffiltext{55}{Technische Universitaet Wien, Wieder Hauptst. 8-10, A-1040 Wienna, Austria; donatowicz@tuwien.ac.at}
\altaffiltext{56}{LATT, Universit\'{e} de Toulouse, CNRS, France; pfouque@ast.obs-mip.fr}
\altaffiltext{57}{Instituto Nacional de Pesquisas Espaciais, Sao
Jose dos Campos, SP, Brazil}
\altaffiltext{58}{NASA Exoplanet Science Institute, Caltech, MS 100-22, 770 south Wilson Avenue, Pasadena, CA 91125, USA; skane@ipac.caltech.edu}
\altaffiltext{59}{Perth Observatory, Walnut Road, Bickley, Perth 6076, WA, Australia;
rmartin,andrew@physics.uwa.edu.au}
\altaffiltext{60}{South African Astronomical Observatory, PO box 9, Observatory 7935, South Africa}
\altaffiltext{61}{Space Telescope Science Institute, 3700 San Martin Drive,
Baltimore, MD 21218, USA}

\altaffiltext{62}{Niels Bohr Institutet, K{\o}benhavns Universitet,
Juliane Maries Vej 30, 2100 K{\o}benhavn {\O}, Denmark}
\altaffiltext{63}{Astronomisches Rechen-Institut, Zentrum f\"{u}r Astronomie der Universit\"{a}t Heidelberg (ZAH),  M\"{o}nchhofstr. 12-14, 69120 Heidelberg, Germany}
\altaffiltext{64}{Institute of Astronomy, University of Zielona G\'ora, Lubuska
 st. 2, 66-265 Zielona G\'ora, Poland}
\altaffiltext{65}{School of Physics, University of Exeter, Stocker Road, Exeter EX4 4QL, United Kingdom}
\altaffiltext{66}{Astrophysics Research Institute, Liverpool John Moores University, Liverpool CH41 1LD, United Kingdom}
\altaffiltext{67}{European Southern Observatory, Karl-Schwarzschild-Stra{\ss}e 2, 85748 Garching bei M\"{u}nchen, Germany}
\altaffiltext{68}{Deutsches SOFIA Institut, Universit\"{a}t Stuttgart,
Pfaffenwaldring 31, 70569 Stuttgart, Germany}
\altaffiltext{69}{SOFIA Science Center, NASA Ames Research Center, Mail Stop N211-3,
Moffett Field CA 94035, USA}
\altaffiltext{70}{Jodrell Bank Centre for Astrophysics, The University of Manchester, Oxford
Road, Manchester M13 9PL, United Kingdom; Eamonn.Kerins@manchester.ac.uk}
\altaffiltext{71}{Max Planck Institute for Solar System Research, Max-Planck-Str. 2,
37191 Katlenburg-Lindau, Germany }
\altaffiltext{72}{Astronomy Unit, School of Mathematical Sciences,
       Queen Mary, University of London, Mile End Road,
       London, E1 4NS, United Kingdom}
\altaffiltext{73}{Centre for Star and Planet Formation, K{\o}benhavns Universitet,
{\O}ster Voldgade 5-7, 1350 K{\o}benhavn {\O}, Denmark}
\altaffiltext{74}{Departamento de Astronom\'{i}a y Astrof\'{i}sica, Pontificia Universidad Cat\'{o}lica de Chile, Santiago, Chile}
\altaffiltext{75}{Universit\`{a} degli Studi di Salerno, Dipartimento di Fisica
``E.R.~Caianiello", Via Ponte Don Melillo, 84085 Fisciano (SA), Italy}
\altaffiltext{76}{INFN, Gruppo Collegato di Salerno, Sezione di Napoli, Italy}
\altaffiltext{77}{Armagh Observatory, College Hill, Armagh, BT61 9DG,
United Kingdom }
\altaffiltext{78}{Institut f\"{u}r Astrophysik, Georg-August-Universit\"{a}t,
Friedrich-Hund-Platz 1, 37077 G\"{o}ttingen, Germany}
\altaffiltext{79}{Bellatrix Astronomical Observatory, Via Madonna de Loco 47,
03023 Ceccano (FR), Italy}
\altaffiltext{80}{Department of Physics, Sharif University of Technology,
 P.~O.\ Box 11155--9161, Tehran, Iran }
\altaffiltext{81}{Institut d'Astrophysique et de G\'{e}ophysique,
All\'{e}e du 6 Ao\^{u}t 17, Sart Tilman, B\^{a}t.\ B5c,
4000 Li\`{e}ge, Belgium}
\altaffiltext{82}{Astrophysics Group, Keele University, Staffordshire, ST5 5BG,
United Kingdom}
\altaffiltext{83}{INAF, Osservatorio Astronomico di Brera, 23806 Merate (LC), Italy}

\begin{abstract}

We present the first measurement of the planet frequency
beyond the ``snow line'', for the planet-to-star mass-ratio
interval $-4.5<\log q< -2$, corresponding to the range of ice giants
to gas giants.  We find 
$$
{d^2N_\pl\over d\log q\, d\log s} = (0.36\pm
0.15)\,{\rm dex}^{-2}
$$ 
at mean mass ratio $q=5\times 10^{-4}$ with no
discernible deviation from a flat (\"Opik's Law) distribution in log
projected separation $s$.  The determination is based on a sample of 6
planets detected from intensive follow-up observations of
%$\sim 10$ 
high-magnification $(A>200)$ microlensing events during
2005-2008. The sampled host stars have a typical mass 
$M_{\rm host}\sim 0.5 M_\odot$,
and detection is
sensitive to planets over a range of planet-star projected separations
$(s_\max^{-1}R_\e,s_\max R_\e)$, where
$R_\e\sim 3.5\,{\rm AU}\,(M_{\rm host}/M_\odot)^{1/2}$
is the Einstein radius and $s_\max \sim (q/10^{-4.3})^{2/3}$.
This corresponds
to deprojected separations roughly 3 times the ``snow line''.
Despite the frenetic nature of these observations, we show
that they have the properties of a ``controlled experiment'', which is
what permits measurement of absolute planet frequency.
High-magnification events are rare, but the survey-plus-followup
high-magnification channel is very efficient: half of all high-mag
events were successfully monitored and half of these yielded planet
detections.  The planet 
frequency derived from microlensing is a factor 7 larger than the
one derived from Doppler studies at factor $\sim 25$ smaller
star-planet separations [i.e., periods 2 -- 2000 days].  
However, this difference is basically
consistent with the gradient derived from Doppler studies
(when extrapolated well beyond the separations from which it
is measured).  This suggests a universal separation distribution
across 2 dex in planet-star separation, 
2 dex in mass ratio, and 0.3 dex in host mass.  Finally, if all planetary
systems were ``analogs'' of the Solar System, our sample would
have yielded 18.2 planets 
(11.4 ``Jupiters'', 6.4 ``Saturns'', 0.3 ``Uranuses'', 0.2 ``Neptunes'') 
including 6.1 systems with 2 or more planet detections.  This compares
to 6 planets including one two-planet system in the actual sample,
implying a first estimate of $1/6$ for the frequency of
solar-like systems.

\end{abstract}


\section{{Introduction}
\label{sec:intro}}

To date, 10 microlensing-planet discoveries have been published,
which permits, at least in principle, a measurement of
planet parameter distribution functions.  Of course, the
size of this sample is small, both absolutely and relative to
the dozens of planets that have been discovered from transit surveys and
the hundreds from Doppler (radial velocity, hereafter RV) surveys.  
However, microlensing probes
a substantially different part of parameter space from these
other methods.  The majority of RV planets, and the
overwhelming majority of transiting planets are believed to
have reached  their present locations, generally well within the
``snow line'', by migrating by a factor $\ga 10$
inward from their birthplace.
By contrast, microlensing planets are generally found beyond
the snow line, where gas giants (analogs of Jupiter and Saturn)
and ice giants (analogs of Uranus and Neptune) are thought to form.
Thus it would be of substantial interest to compare the
properties of microlensing planets, which have not suffered major
migrations, to other planets that have.  

In the present incarnations of microlensing experiments, planet
detection occurs through several different routes, with various
degrees of ``human intervention''.  At one extreme, planets can
be detected simply on the basis of survey data, without even
the knowledge that a planet was  present until after the
event is over.  This is the plan for so-called second generation
microlensing planet surveys, which will permit rigorous,
particle-physics-like objective analysis of a ``controlled
experiment''.  In fact, \citet{tsapras03} and \citet{snodgrass04}
have already carried out such analyses of first-generation survey-only 
data. However, to date only one secure microlensing
planet  has been discovered from survey-only data, 
MOA-2007-BLG-192Lb \citep{mb07192}.

Rather, most microlensing planet detections have taken place through
a complex interplay of survey and followup observations.
Only a year after the first microlensing events
were discovered, the MACHO collaboration issued an IAU circular
urging follow-up observations of a microlensing event, and
the OGLE collaboration initiated 
its Early Warning System (EWS) \citep{ews1,ews2}, which regularly issued
``alerts'' of ongoing microlensing events, usually well before peak,
as soon as they were reliably detected.
EWS, together with a similar program soon initiated by the
MACHO collaboration \citep{alcock96} with their much larger camera, 
enabled the formation in 1995 of the
first microlensing follow-up teams, PLANET \citep{albrow98} and
MACHO/GMAN \citep{gman}, and soon thereafter, MPS \citep{rhie99}.  
The EROS team issued only a few alerts, but one of these led
to the first mass measurement of an unseen object \citep{an02},
and the first spatially resolved high-resolution spectrum of another
star \citep{castro01}.
OGLE-II inaugurated wide-field observations as well in 1998.
In order to cover large areas of the
sky (even with wide-field cameras), the survey teams would generally 
obtain only $\sim 1$ point per
night per field.  Since typical planetary deviations 
from ``normal'' (point lens) microlensing light curves
last only of order a day or less, this was not generally adequate to
detect a planet.  Hence,  \citet{gouldloeb92} already advocated 
the formation of follow-up teams that would choose 
several favorable events to monitor more frequently, using
telescopes on several continents to permit 24-hour
coverage.  In 2000, MACHO ceased operations, but a new
survey group, MOA, had already begun survey observations.

Once this synergy between survey and follow-up teams was established, 
it evolved quickly on both sides.  Both types of teams developed
the capacity for ``internal alerts'', whereby real-time photometry
was quickly analyzed to find hints of an anomaly.  
If these hints
were regarded as sufficiently interesting, they would trigger
additional observations by the team.  The first such
alert was by MACHO/GMAN in 1996 and several followed the next year
from PLANET.   In 2003, OGLE developed the Early EWS,
which automatically alerted the observer to possible anomalies,
who then would make additional observations and, if these were
confirming, publicize them to the community.  OGLE also
pioneered making their data publicly available, which greatly
facilitated follow-up work.  These developments generally evolved  
into a system of mutual alerts, open transfer of data between teams, and
active ongoing email discussion of developing events.

The first secure microlensing planet, OGLE-2003-BLG-235Lb \citep{ob03235}
was discovered by means of such an internal alert.  The MOA
team noticed deviations in their data, and initiated intensive
follow-up of this event.  In retrospect, one can see that even
without this internal alert, the combined OGLE and MOA data would
have been sufficient to show that the star had a companion.  However,
the normal survey data would not have permitted one to tell whether
this companion was a planet or a low-mass star (or possibly a
brown dwarf).  

Detections of this type are at the opposite extreme from the
pure-survey detection, which can be modeled as a controlled 
experiment.  Without understanding the efficiency at which
the observers issue their real time alerts, one cannot
measure the absolute rate of planets from observations that
are triggered by the presence of the planet itself.
Another planet, OGLE-2007-BLG-368Lb \citep{ob07368} falls partly
into this category. In this case, the alert was triggered
by ARTEMIS \citep{dominik08}, whose ultimate
goal is to communicate such alerts directly to the followup telescopes
without human interference (and so bring such alerts into the
fold of ``rigorous controlled experiments''), although at this stage
ARTEMIS alerts were still vetted by humans. A further
anomaly was soon detected by the MOA observer.  But even without
followup observations triggered by these alerts, 
this companion probably could have
been constrained to be planetary, although with larger errors,
so this detection could still be integrated into the ``controlled
experiment'' framework even without trying to model the alert
process.

However, if follow-up observations are carried out without being
significantly influenced by the possible presence of a planet,
then these also can be treated as a controlled experiment,
and absolute rates can be derived using either the method of
\citet{gaudi00} or of \citet{rhie00}. 
Such an analysis was carried out for the first 43 events
monitored by the PLANET Collaboration \citep{albrow01,gaudi02}.
In particular, since no planets were detected (and so only
upper limits obtained), there was no possibility of
recognition of a possible planet playing any role in the
observation strategy.  But, as emphasized above, once
planets started to be discovered, the situation became more
nuanced.  

One path toward obtaining followup observations that are 
triggered on potentially planetary anomalies, without
compromising the ``controlled experiment'' ideal, is to
establish robotic triggering programs that communicate
directly to robotic telescopes.  This has been the goal
of the RoboNet Collaboration since its inception in 2004
\citep{tsapras09}.
Algorithms for detecting anomalies in robotically acquired
real-time data \citep{dominik07,dominik08} and directing robotic
telescopes \citep{horne09} have been devised and are being
further developed, and
will expand their scope as the Las Cumbres Observatory Global
Telescope Network (LCOGT) itself expands.  But to date, the
most effective RoboNET observations have been ``hand triggered''
(e.g., for OGLE-2007-BLG-349Lb).

In spite of this nuanced situation,
\citet{ob07368} were nevertheless able to extract
relative frequencies of planets from the ensemble
of all 10 microlensing planets.  They argued that each of
these planet detections was characterized by a sensitivity
$g(q)\propto q^m$, where $q$ is the planet/star mass ratio.
While $m$ varies from event to event, they argued that $m=0.6\pm 0.1$
was an appropriate average value.  They then fit the ensemble
of detections to a power-law mass distribution 
$f(q)d\log q\propto q^n d\log q$ and found $n=-0.68\pm 0.20$.
But they did not attempt
to extract absolute frequencies from their sample.

Here, we analyze an important subclass of microlensing
planet searches: high-magnification events that are intensively
monitored over the peak.  The observations
of these events are always frenetic, sometimes even comical,
so it may seem surprising that they nevertheless constitute 
a ``controlled experiment'', or very nearly so, and hence are
subject to rigorous analysis.

More than a decade ago \citet{griestsafi} pointed out that
high-mag events are much more sensitive to planets than their
far more common cousins, the low-magnification events.  The
reason is fairly simple.  For point-lens/point-source microlensing, the
magnification is very nearly $A\rightarrow u^{-1}$ for $u\la 1/3$, where 
$u$ is the source-lens separation normalized to the Einstein radius.  
``High-magnification''
($A_\max\gg 1$) therefore implies impact parameter $u_0=A_\max^{-1}\ll 1$,
i.e., that the source is projected very close to the lensing star.
If the lensing star has a planet, the planet's gravitational field
will perturb the otherwise circularly symmetric gravitational 
field of the star,
and so induce a tiny ``central caustic'' (contour of infinite magnification)
around it.  If the source passes over or near this caustic, the
lightcurve is perturbed, revealing the planet's presence.  The key
point is that the planet can be at a broad range of separations and
at virtually any angle relative to the source trajectory and still
create a caustic that will perturb the light curve.

However, following this seminal paper, microlensing follow-up 
monitoring groups continued to focus primarily on garden variety microlensing
events, which are only sensitive to so-called ``planetary caustics''.
These are formed by the action of the primary-star gravitational
field on the planet gravitational field, and as such are bigger
and so have larger cross sections than central caustics.  
At first sight, this makes them
more favorable targets, but this conclusion only holds if one has
unlimited telescope time for monitoring.  Then one would monitor
all available events (and hence primarily low-mag events) and would
find most planets in these events (just because the caustics have a larger
cross section).  But if observing resources are limited, then
one should focus these on high-mag events because these can
be predicted (at least in principle) from the pre-peak part of the
light curve and have individually higher sensitivity to planets.
By contrast, there is no way to predict that a source is approaching
a planetary caustic.

Nevertheless, within the context of continued focus on planetary caustics
in normal events, there were significant efforts to take advantage of 
high-mag events as well.  In 1998 MPS issued the first high-mag alert
for MACHO-98-BLG-35, which received an enthusiastic response from
the MOA group, leading to the possible detection of a planet
\citep{rhie00,bond02b}, but with too-low significance to be confident.
 From the inception of its survey, MOA made real-time alerts of
high magnification events a priority and attempted to organize
follow-up from other continents, with 10 such alerts the first year
\citep{bond02a}.  The most spectacular success of this program was
MOA-2003-BLG-32/OGLE-2003-BLG-219, which was densely sampled over
its $A_\max=520$ peak by the Wise observatory in Israel after
such an alert, and which yielded the best upper limits on planetary
companions to a lens to that date \citep{abe04}.  Theoretical work
was also done to optimize observations of high-mag events for
planet sensitivity \citep{rattenbury02}.

When the Microlensing Follow Up Network ($\mu$FUN) began operations
in 2001, it followed the already-established model of follow-up
observations, which did include high-mag events \citep{ob03423},
but did not emphasize them.  However,
three things happened to change its orientation toward concentrating
on high-mag events.  First, the OGLE-III survey came on line
in 2002 with a discovery rate of 350 events per year, moving
up to 600 events per year in 2004.  This compared with 40--80
events per year discovered by OGLE-II in 1998--2000.  
The number of events alerted per year has
a direct impact on whether or not one is in the regime of ``limited'' or
``unlimited'' resources.  Before 2002,
if one restricted oneself to high-mag events, one would mostly
be sitting on one's hands.  For example, when \citet{albrow01} and
\citet{gaudi02} analyzed 5 years of PLANET collaboration data,
they reported only two events with $A_\max>100$.  OGLE-III 
dramatically changed that situation.  More recently, the MOA
collaboration inaugurated MOA-II (2007 first full season), which
has had the net effect of increasing the total rate of reported events
by about 50\%.

Second, $\mu$FUN began attracting the intrinsically ``limited''
observing resources of amateur astronomers.  These contrast with
the larger dedicated professional observatories in two key ways.
First, the observers generally cannot observe all night, every
night, or they will be unable to keep their day jobs.  Second,
the smaller apertures of their telescopes restrict them to
relatively brighter targets.  Both ``limitations'' naturally
drive amateurs to high-mag events, which have a bigger chance
of science payoff and are brighter (because highly magnified).  
Moreover, there is one crucial dimension in which amateurs are not
limited: they have completely
free and almost instantaneous access to their telescopes at any time.
Thus, while dedicated follow-up telescopes are typically operated
only in the 3-4 month core of the season (when microlensing targets
are observable for at least half the night), amateurs can react
to alerts deep into the wings of the season, close to doubling
the number of high-mag events that can be monitored.
In 2004, one amateur began observing $\mu$FUN targets on her
own initiative.  She requested regular alerts and began organizing
other amateurs to join in, who in turn self-organized a network.  
About half of the $\mu$FUN authors of this paper are amateurs.

Third, $\mu$FUN had to become aware of this changed situation, i.e.,
that it had moved from the domain of unlimited to limited resources.
This transition was partly aided by the fact that $\mu$FUN access
to two of its professional telescopes (Wise and SMARTS CTIO) was limited
by their being shared resources.  But throughout 2003-2004, $\mu$FUN
``straddled two horses'', focusing on high-mag events when available,
but trying to keep to the old planetary-caustic strategy most of
the time.  Preparations for the 2005 season were significantly
influenced by preliminary work (ultimately, \citealt{ob04343}) showing
that unless high-mag events were intensively monitored over their
peak, much of their sensitivity to planets is compromised.

Then in April 2005, $\mu$FUN intensively followed the
(by those days' standards) high-mag event OGLE-2005-BLG-071, which
resulted in the detection of the second microlensing planet
\citep{ob05071}.  This detection led to $\mu$FUN consciously
changing its orientation, procedures, recruitment, etc., with the
aim of focusing primarily on high-mag events.  

 From 2005 onwards,
considerable effort has gone into identifying potential high-mag events,
and in some cases obtaining additional data to improve the prediction
of $A_\max$. If an event is deemed a plausible high-mag candidate,
then observers are notified by email, without necessarily being urged
to observe, just to put them on alert.  Once high magnification
seems probable, observations are requested with various degrees of
urgency.  The urgency is conditioned by the fact that peak sensitivity
is usually less than a day (the normal human cycle time) but more than a few 
hours (so requiring observations from multiple continents).
Many factors enter into the quality of the final
light curve, including weather conditions on 6 continents plus 
Pacific Islands, observer availability, communication problems, etc.
Indeed, it is difficult to convey the level of chaos during one of
these events.

Despite (and also because of) this chaos, 
the resulting data stream generally retains
the character of a ``controlled experiment''.
In the ``ideal incarnation'' of this
search mode, the event is recognized (with greater or
lesser certainty) to be approaching a high-magnification peak,
and an alert is issued to interested observers urging them
to observe it intensively over the predicted peak.  The
observations take place regardless of whether the planet is
present or not.  The very chaos, remoteness of observing locations
and communication problems make it difficult to gain knowledge
of planetary perturbations until after the key observations
are over.  In Section \ref{sec:data}, we discuss how well
the real events conform to this ideal, but assuming for the
moment that they do, the ensemble of high-magnification events
with and without detections can be analyzed in exactly the same
way as can be done for the ``pure survey'' mode.
In Section \ref{sec:analytic}, we derive an analytic expression
for the sensitivity of an ensemble of high magnification
events monitored densely over their peak, and
in Section \ref{sec:numerical}, we calculate numerical sensitivities
and use these to measure the frequency of ice giants and gas giants beyond
the snow line.  Finally, in Section \ref{sec:discuss},
we discuss some implications of our results.

{\section{Selection Criteria and Data}
\label{sec:data}}

We begin by designing criteria for selecting events 
and planets to be included
in the analysis that enable the
detections and non-detections to be analyzed on the same
footing.  This is essential to the goal of defining a
data sample that can be treated as a ``controlled experiment''.

{\subsection{Selection Criteria for Events}
\label{sec:eventselection}}

\hbn E1) High-cadence ($<10\,$min) $\mu$FUN data over some portion of the peak
\hbn E2) High-cadence data covering at least one wing of the peak,
        $|t-t_0|<t_{\rm eff}$, with no major gaps
\hbn E3) High signal-to-noise ratio (S/N) data ($\sum_i \sigma_i^{-2}>50000$)
covering the other wing, where 
the $\sigma_i$ are the data-point errors in magnitudes
\hbn E4) Light-curve is not dominated by binary (non-planetary) features

We define $t_{\rm eff}\equiv u_0 t_\e$ for point-lens/point-source
events, where $t_\e$ is the Einstein crossing time.  For 
point-lens events that suffer finite-source effects, we generalize
this definition to the time interval during which the magnification
is within $\sqrt{2}$ of the peak.  And for planetary events, we further
generalize it to the time interval that the magnification would have
been within  $\sqrt{2}$ of the peak if the lensing star had lacked planets.

We now justify these criteria.

In addition to reflecting the fact that we are summarizing $\mu$FUN work,
criterion (E1) ensures that we can rigorously review the available
data on $\sim 3000$ events discovered during 2005-2008, and reduce
them to a manageable subset of ``only'' 315 that can be investigated
using $\mu$FUN files.  These 315 were then quickly pared down to a few
dozen events that are consistent with high-magnification and actually
meet criterion (E1).

Criteria (E2) and (E3) should be considered together.  Our underlying
requirement is to have enough coverage of the event so that
planetary deviations that give rise to a $\Delta\chi^2=500$
deviation (for the rereduced data set), 
have a high probability of yielding a unique scientific
interpretation.  See criterion (P2), below.
These two criteria are basically derived from the experience analyzing 
the event
OGLE-2005-BLG-169 \citep{ob05169}, which does somewhat better than
barely satisfy both.  The analysis of this event was already
fairly difficult because of the multiple $\chi^2$ minima and would
have been quite degenerate if, for example, it lacked high S/N data on
the rising side.  Hence it would not have led to a publishable planet
with a reasonably well-defined mass ratio.

Criterion (E4) is adopted for two reasons.  First, in contrast to
point-lens events, most binary events are not modeled with sufficient
precision to measure the $u_0$ parameter well enough to
construct a well-defined sample of events with maximum magnification
greater than some threshold $A_\max$.  Second, the problem of
detecting planets in the presence of light-curve features dominated
by a binary is not well-understood.  Hence, the results we derive
here really apply to stars not giving rise to strong binary features,
which excludes roughly 3--6\% of all stars \citep{macho-binary,ogle-binary}.

Table \ref{tab:events} lists all events from 2005-2008 that
satisfy these four 
criteria and that had peak magnifications $A_{\rm max}> 100$.
Column 1 is the event name,  column 2 is the maximum magnification,
column 3 is the time of closest approach between the source and lens
$t_0$, column 4 is the Einstein timescale $t_\e$, column 5 gives the 
mass of the lens star
for cases that it is known, and column 6 gives the method by which it is
derived.  These lens masses and methods are discussed in Section 
\ref{sec:hostmass}.
The events are listed in inverse order of $A_\max$.

Figure \ref{fig:amax} shows the cumulative distribution $A_\max^{-1}$
for the 15 events in Table \ref{tab:events}.
It displays a clear break at
$A_{\rm max}=200$.  Below this value, the distribution is uniform in
$A_{\rm max}^{-1}$, which is the expected behavior for a
complete sample (ignoring finite source effects), i.e.,
uniform in $u_0$, which is the impact parameter in units
of the Einstein radius.  The dashed line, with
slope of $d N_\ev/d A_\max^{-1}=2600$ is a good match to the
$A_{\rm max}>200$ data.  \citet{cohen10} found such uniformity
for the underlying sample of OGLE events in 2008, with a slope
of $d N_\ev/d u_0 = 1080$ for $u_0<0.05$.  From this comparison,
we learn two things.  First, $\mu$FUN was aggressive enough to 
achieve a uniform subsample only for events with $A_{\rm max}>200$.
Second $\mu$FUN was able to intensively
monitor half of all events in the $A_{\rm max}>200$ subsample.
That is, assuming that OGLE found similar numbers of high-mag events in
2005-2007, and accounting for the fact that MOA found 50\% more
events (not found by OGLE) in 2007-2008, the full sample of high-mag events
was about $d N_\ev/d u_0 \sim 5\times 1080 =5400$, of which $\mu$FUN
effectively monitored about 48\%.

Although it may not be immediately obvious, Figure \ref{fig:amax}
implies that we {\it must} impose a fifth criterion.

\hbn E5) $A_\max > 200$

Figure \ref{fig:amax} 
demonstrates that $\mu$FUN was substantially less enthusiastic
about events $A_\max<200$ than $A_\max>200$, whether because it
simply did not act on events known in advance to be in the former
category or just became less enthusiastic about observations
once these events were recognized near peak not to be extremely magnified.  
This bias is a natural consequence
of $\mu$FUN's limited observing resources (as discussed
in Section \ref{sec:intro}):  there are 4 times as many events with
$A_\max >50$ as $A_\max>200$ and their duration of peak, 
$2\,t_{\rm eff} \sim 2\,t_\e A_\max^{-1}$ lasts 4 times as long, so
16 times more observing resources would be required to follow them all.
Hence, if an event proved midway to be one that the objective evidence
demonstrates $\mu$FUN cared less about, there would be a tendency on the
part of observers to slacken efforts (whether or not the internal
alert was officially called off).  Then the event would have less chance
of meeting the selection criteria.  But if a planet were detected
during the peak observations 
(and there is a greater chance of recognizing a planet in real time 
for lower $A_\max$
because the peak lasts longer) then observations would not slacken,
but rather intensify.  Since this bias cannot be rigorously quantified,
planets and non-detections from $A_\max<200$ events must both be 
excluded from the sample.

{\subsection{Selection Criteria for Planets}
\label{sec:planetselection}}

\hbn P1) Planet must be discovered in an event that satisfies (E1)-(E5)
\hbn P2) Planetary fit yields improvement $\Delta\chi^2>500$
\hbn P3) Planet-star mass ratio $q$ must lie in the range
\begin{equation}
q_- < q < q_+,
\qquad q_- = 10^{-4.5}, 
\qquad q_- = 10^{-2}.
\label{eqn:massrange}
\end{equation}

Criterion (P1) is self-evident but is stated explicitly for
completeness and emphasis.  Criterion (P2) may appear at first
sight somewhat draconian, but it is realistic.  To explain this,
we first note that among the six planets listed in
in Table \ref{tab:parameters}, the ``weakest'' detection is
$\Delta\chi^2=880$, which is for MOA-2008-BLG-310Lb
\citep{mb07400}.  Now, it is certainly possible to recognize
systematic residuals from a point-lens fits ``by eye'' 
at a much lower level, even $\Delta\chi^2=100$.  Indeed,
\citet{ob07050} argued that no systematic residuals were
present in the fit to OGLE=2007-BLG-050 at a much lower
level, $\Delta\chi^2=60$.  But if such deviations had been
observed in an event, this would not have necessarily 
enabled discovery of a planet, where ``discovery'' here means
``publication''.  First, $\Delta\chi^2=100$ in the final,
rereduced and carefully cleaned data implies something like
$\Delta\chi^2=50$ in the standard pipeline data, and systematic
deviations due to a planet at this level would probably not be 
recognized as significant,
i.e., clearly distinguishable from systematics that appear in
dozens of other events, and which just reflect instrumental,
weather, or data-reduction problems.  But more to the point,
even if the unrereduced-data were $\Delta\chi^2=100$, triggering
strenuous efforts to clean and rereduce the dataset, resulting
in, say, a $\Delta\chi^2=200$ improvement,  it is far from clear
that this deviation (even if strongly believed to be real)
would lead to a publishable planet detection.  This is because,
in addition to obtaining an acceptable fit to the data,
such a paper would have to demonstrate that there could
be no acceptable fits to the data to non-planetary solutions.
We have already designed criteria (E2) and (E3) to eliminate
those events for which very high $\Delta\chi^2$ is possible
without leading to a unique interpretation (due to
incomplete coverage of the deviation).  But we still must
set the threshold high enough so that if an anomalous event survives
criteria (E2), (E3), and (P2), it has a small chance of being
ambiguous in its interpretation.  Our best estimate of this,
from experience fitting events, is $\Delta\chi^2 = 500$.
However, we regard other values in the range 350-700 as also
being plausible candidates for this threshold.  We will show
in Section \ref{sec:numerical} that our basic conclusions are
robust to changes within this range.

The upper boundary in Equation (\ref{eqn:massrange}),
Criterion (P3), is necessary
because at high mass ratios $q$, one cannot be confident that
the event will not be rejected (consciously or unconsciously)
as a ``brown dwarf'' or ``low-mass-star'', and therefore not
be monitored as intensively as it might be [and so not pass
criteria (E2) and (E3).]  To illustrate this, we review how
OGLE-2008-BLG-513Lb (Yee et al., in preparation) was ``almost rejected''
as a binary, even though it is probably a planet.  This
event has a large, strong, resonant caustic, that was
initially mistaken for a binary.  During the long
intra-caustic period, it was realized that the companion
might be a planet, and that intensive observations of the
caustic exit would be necessary to resolve this question.
Such observations  were obtained, and from these we know
that the impact parameter was $u_0\sim 0.027$,  so $A_\max\sim 37$,
which means that the event fails criterion (E5) and so is
excluded from our sample.  But if these data had not been obtained,
then $u_0$ for this event would not be known, and it would not
be known whether the event was in the sample or not, and if
it were, whether the companion was a planet or not.

Why does this example then not just prove that the whole concept
of ``controlled experiment'' is unviable?  The answer is
given by Figures \ref{fig:ob07050} and \ref{fig:ob08279}, below.
One sees from these that at high mass ratios, $q\sim 10^{-2}$,
there is an extremely wide range of $s$ for which the
planet is ``detectable''.  Only a small fraction of
these ``detectable'' events have $s\sim 1$, which produce
large resonant caustics that might be mistaken for binaries.
Hence, while in principle some of these ``detections'' might
be lost to this confusion, the great majority would not
cause any confusion.  The reason that planets like OB08513Lb
make their way into the detections at all, despite their
relative rarity, is that the caustics are so large that
they are detectable over a wide range of $u_0$, only a 
small fraction of which would pass the ``high-mag'' criteria
of Section \ref{sec:eventselection}.  Nevertheless, as $q$
grows, this potential problem grows with it. We adopt
$\log q_+ = -2$, but recognize that values ranging from $-2.3$
to $-1.8$ might also have been plausible choices.

The lower boundary $q_- = 10^{-4.5}$ is established because of 
concerns of the real ``detectability'' of low mass planets in
the presence of higher mass planets in the same system.
In the method of \citet{rhie00}, which we employ in Section
\ref{sec:fullsample}, planet sensitivity 
is determined by fitting simulated star-planet light curves that
are constructed to have the same error properties as the actual
data, to point-lens models.  If the best such model increases
$\chi^2$ by more than a given threshold (say $\Delta\chi^2>500$), then the
planet is said to be detectable.  Since this method directly
mimics the process of planet detection for single-planet
systems, it is a good way to characterize the detectability
of such systems.  But high-magnification events are particularly
sensitive to {\it multiple planets} \citep{gaudi98} and why should this
approach tell us anything about the
detectability of a second planet in a system already containing
one planet?  As first shown by \citet{bozza99}, the net perturbations of such 
two-planet high-magnification lightcurves usually ``factor'' 
into the sum of perturbations induced by each planet separately.
For example, the only published two-planet system has this
property \citep{ob06109,bennett10}.  Of course, the factoring
is not perfect, but in this real case (and in many simulated
cases), once the dominant-planet perturbation is removed, the
secondary perturbation is easily recognized, leading to an
excellent starting point for a combined fit to both planets
simultaneously.  For reasonably comparable planet mass ratios, 
the only exception to this is if the planet-star axes are
closely (within $\la 20^\circ$) aligned \citep{rattenbury02,han05}.  
In this case, the single-planet fit still
fails, but the residuals to this fit are not easily recognizable.
While such difficulties might impede recognition of the second
planet, the required alignment is so close, that such cases would
be a small minority of two planet systems.

However, this factoring has only been studied in detail for
planets with relatively comparable masses \citep{han05}.
The situation may not be as simple when the mass ratio of the
two planets is extreme.  Based on analysis of the events listed
in Table \ref{tab:parameters}, below, we cannot be confident of excluding
all ``second planets'' with $q<10^{-5}$.  To be ``conservative'', we
have moved the boundary to $q_-=10^{-4.5}$.

Because both $q_-$ and $q_+$ have some uncertainty, we must ask
how robust our conclusions are to changes in these parameters
within a reasonable range.  We show in Section \ref{sec:numerical}
that our basic conclusions about planet frequency are not seriously
affected by uncertainty in $q_-$ and $q_+$.  However, these
uncertainties will prevent us from deriving a slope of the mass-ratio
function.

Six planets satisfy criteria (P1)-(P3).  
Table \ref{tab:parameters} displays their characteristics.
Columns 2--5 are the parameters $(u_0,\rho,q,|\log s|)$
measured from the event, where $\rho$ is the source radius
in units of the angular Einstein radius $\theta_\e$.
In several cases, there is an
unresolved $s\leftrightarrow s^{-1}$ degeneracy, which is
irrelevant to the current study, so we just display the
absolute value of the log.  The final column is the
maximum detectable value of $|\log s|$ according to 
calculations reported in Section \ref{sec:fullsample}.

{\subsection{Were Discovery Observation Cadences Really Independent of the
Planet? }
\label{sec:independent}}

Were the observations that led
to the discovery of the six planets in fact carried out
independent of the presence of the planet?
For three of these, 
OGLE-2005-BLG-169Lb \citep{ob05169},
MOA-2007-BLG-400Lb \citep{mb07400}, and
MOA-2008-BLG-310Lb \citep{mb08310}, the planet was not
recognized until after the event had returned to baseline.
OGLE-2007-BLG-349 (Dong et al., in preparation) was recognized
to have a significant deviation possibly due to a planet based
on observations in Chile, 36 hours after the call for intensive
observations based on its high-mag trajectory, and roughly 7 hours 
after observations had begun in South Africa.  While it is true that reports
of this potential planet heightened excitement, and may perhaps have
increased the commitment of observers to get observations, the
density of observations (from 4 continents plus Oceania) did not 
qualitatively change from before till after the potentially-planetary
anomaly was recognized.  This is the only one of the six planets
in our sample that is not yet published.  The reason is that
the system contains a third body, which has proven difficult
to fully characterize.  However, the characteristics of
OGLE-2007-BLG-349Lb are very well established, and the third
body is certainly not in the mass range being probed in the current
analysis.  Hence we feel confident including this planet in the sample.

The two planets OGLE-2006-BLG-109Lb,c \citep{ob06109,bennett10} require
closer examination. OGLE-2006-BLG-109  was recognized to be
an interesting event almost 10 days prior to peak due to
detection of what turned out to be the resonant caustic of
OGLE-2006-BLG-109Lc, the Saturn mass-ratio planet in this system.  
This anomaly did indeed
trigger some additional observations, which did help characterize
this planet.  But a review of email communications that
initiated follow-up observations during the
event shows that far more intensive observations 
were triggered several days later, after the event had appeared to
return to ``normal'' (point-lens-like) microlensing, 
exactly by its high-mag trajectory.  Although
these emails remark on the possible presence of a planet,
they place primary emphasis on this being an otherwise ``normal''
microlensing event that was reaching extreme magnification.
(Note that the appearance of extreme magnification was not itself
an artifact of the presence of planets(s), but was simply due
to low source-lens impact parameter.)\ \ 
It was the intensive observations from New Zealand,
triggered by these emails,
that captured the ``central structure'' of the caustic due to
OGLE-2006-BLG-109c.  These would have enabled basic characterization
of this planet even without the flurry of followup observations
10 days earlier.  Moreover, it was the same email that triggered
intensive observations 
from two widely separated locations
(Israel and Chile), that enabled detection of the ``central caustic'' 
due to OGLE-2006-BLG-109Lb, the Jupiter mass-ratio planet.  
Thus, all detections were in 
reasonable accord with the ``controlled experiment'' ideal.

{\section{Analytic Treatment}
\label{sec:analytic}}

{\subsection{Triangle Diagrams}
\label{sec:triangle}

\citet{ob07050}  and \citet{ob08279} recently analyzed the
sensitivity to planets of two events in our sample from
Table \ref{tab:events}, OGLE-2007-BLG-050 and OGLE-2008-BLG-279 
respectively.  Figures \ref{fig:ob07050} and \ref{fig:ob08279}
are versions of their results, but with the detection threshold
used in this paper, $\Delta\chi^2=500$.
In contrast to the wide range of detection-sensitivity morphologies
shown in Figure 8 \citet{gaudi02}, both of these diagrams have a 
simple triangular appearance, which is basically described by
a two-parameter equation,
\begin{equation}
|\log s|_\max(q) = \eta\log{q\over q_\min},
\label{eqn:triangle}
\end{equation}
where $q$ is the planet/star mass ratio, $s$ is the 
planet-star projected separation in units of the Einstein radius,
$\eta$ is the slope of the triangle,
and $q_\min$ defines the ``bottom'' of the triangle.  
Planets lying inside the triangle are detected with 100\%
efficiency and those lying outside are undetectable.  The
boundary region is quite narrow.  In principle, 
planet detection is a function not only of $(s,q)$, but also
$\alpha$, the angle of the source-lens trajectory relative
to the planet-star axis.  The narrowness of the boundary reflects
that detection is almost independent of $\alpha$.  See
\citet{ob07050}  and \citet{ob08279} for concrete illustrations.

It is also striking that the slopes $\eta\sim 0.32$  and $\eta\sim 0.35$ 
are nearly the same
for the two diagrams, leading to the conjecture that $\eta$
is very nearly constant for well-monitored high-magnification
events.  Indeed, of the 43 events analyzed by \citet{albrow01} and
\cite{gaudi02}, two are relatively high-mag ($A_\max>100$) and
both have the same triangular appearance and very similar slope
$\eta$, as does the extreme $A_\max=3000$ event OGLE-BLG-2004-343
with simulated coverage analyzed by \citet{ob04343}. If truly generic,
this would mean that high-mag event sensitivities have an extremely
simple triangular form characterized by a single parameter, $q_\min$.

Moreover, while $q_\min$, i.e. the depth of the triangles in
Figures \ref{fig:ob07050} and \ref{fig:ob08279}, obviously depends
on the intensity, quality, and uniformity of coverage, one
expects the fundamental scaling to be
\begin{equation}
q_\min = \xi A_\max^{-1},
\label{eqn:qmin_u0}
\end{equation}
where $\xi$ is a parameter that depends on the data quality, etc.
The reason for this expected scaling is that the size of the central
caustic is proportional to $q$,  and $A_\max^{-1}$ measures
how closely the source probes the center, which is roughly the
maximum of the impact parameter $u_0$ and the source size $\rho$
\citep{hankim09,ob07050}

Hence, armed with an empirical estimate of $\xi$, one can
quickly gage the sensitivity of one event or an ensemble of
events to planets, which is quite useful both to guide observations
and as a check on
``black-box'' simulations of event sensitivity.  Indeed as we
will show below, one can approximately ``read off'' the frequency
of planets by just counting the number detected and the number of
high-mag events surveyed.  

Based on this handful of published analyses, we estimated $\xi\sim 1/70$
for a $\Delta\chi^2=500$ threshold.  Since these analyses (naturally)
focused on events with better-than-average 
coverage, we estimate that $\xi\sim 1/50$
is more appropriate for a sample such as ours.

{\subsection{Analytic Estimate From Triangles}
\label{sec:estimate}

We now assume that planets are distributed uniformly in $\log s$
[\"Opik's law \citep{opik24}] in the neighborhood of the Einstein 
ring\footnote{
If in fact planets are distributed $dN/d\log s\propto s^p$,
then Eq.\ (\ref{eqn:prob}) is in error by $\sinh x/x$ with 
$x=\eta p\ln q/ q_\min$.  For $\eta=0.32$, $p=0.4$ and $q/q_\min = 100$,
this is still only a factor 1.06.
}.  
We will show further
below that this assumption is consistent with current microlensing
data.  Then, assuming $\eta=0.32$ to be universal, we have
\begin{equation}
P_i(q)d\log q = 2\eta f_i(q)\log{q\over q_{\min,i}}
\Theta({q- q_{\min,i}}) d\log q
\label{eqn:prob}
\end{equation}
where $f(q)$ is the number of planets per dex of projected separation 
per dex of mass ratio, and
$\Theta$ is a step function.  The expected number of planets
detected in high-mag events is then just the sum of Equation (\ref{eqn:prob})
over all high-mag events with good coverage
\begin{equation}
\label{eqn:nofq0}
{d N_\pl\over d\log q} = \sum_{i=1}^{N_\ev} P_i(q).
\end{equation}
Figure \ref{fig:amax} demonstrates that
the $\mu$FUN sample is uniform for
$0<A_{\rm max}^{-1}<\epsilon$, where $\epsilon = 0.005$.
Hence, we can turn this sum into an integral,
$$
{d N_\pl\over d\log q} = \sum_{i=1}^{N_\ev} P_i(q) \rightarrow 
{N_\ev\over\epsilon}\int_0^\epsilon P_{A_\max}(q) d A_\max^{-1}
=2\eta f(q){N_\ev\over\epsilon}\int_0^\epsilon 
\log[q/(\xi/A_\max)]\Theta(q-\xi/A_\max)])d A_\max^{-1}
$$
\begin{equation}
\label{eqn:nofq1}
={2\eta\over \xi}f(q){N_\ev\over\epsilon}\int_0^{\xi\epsilon} 
\log(q/q_\min)\Theta(q-q_\min)d q_\min
={2\eta\over \xi \ln 10}
f(q){N_\ev\over\epsilon}\int_0^{\min(q,\xi\epsilon)} 
\ln(q/q_\min)d q_\min,
\end{equation}
which may be evaluated,
\begin{equation}
\label{eqn:nofq2}
{d N_\pl\over d\log q}={2\eta N_\ev\over \ln 10}g(q)f(q)
\end{equation}
\begin{equation}
\label{eqn:gofq}
g(q)= {q\over q_\thr}
\quad (q<\xi\epsilon);
\qquad
g(q) = 1 + \ln{q\over q_\thr}
\qquad (q>\xi\epsilon)
\end{equation}
where $q_\thr \equiv \xi\epsilon$.
Below this threshold, detection efficiency falls linearly with $q$.
Above the threshold, it rises logarithmically with $q$.
Note that the appearance of ``$\ln 10$'' in these formulae is
an artifact of our having chosen to express the density of
planets in units of dex of separation, rather than the ``natural
unit'' of an $e$-folding.

Both the break at $q_\thr$ in the normalized survey sensitivity $g(q)$
and the functional forms of $g(q)$ on either side of this
break are easily understood from the triangular form of the
individual-event sensitivity diagrams.  For $q>q_\thr$, all 
$N_\ev$ events contribute sensitivity.  If we compare two mass
ratios, $\log q$ and $\log q+d\log q$, the latter is sensitive to
a log-separation interval on the triangle that is larger
by exactly $2\eta d\log q$ for {\it each individual event}, so
the sensitivity of the ensemble of events is simply 
$2\eta\log q + {\rm const}$.  On the other hand, for 
$q<q_\thr$, only a fraction $q/q_\thr$ of the events contribute,
which breaks the logarithmic form of $g(q)$.

Note that our estimate $\xi = 1/50$ implies that our
$A_\max< \epsilon^{-1}=200$ survey has
\begin{equation}
\label{eqn:qthr}
q_\thr = \epsilon\xi = 10^{-4},
\end{equation}
i.e., twice the mass ratio of Neptune.  We discuss the implications
of this threshold in Section~\ref {sec:discuss}.

{\section{Numerical Evaluation}
\label{sec:numerical}}

{\subsection{Sensitivities for the Full Sample}
\label{sec:fullsample}}

To more accurately determine the sensitivity of our survey and
to infer the frequency of planets, we carry out a detailed 
sensitivity analysis of all 13 of the $A_\max>200$ events in
Table \ref{tab:events} (except the two that were already
done).  We use the method of \citet{rhie00} (outlined in
Section \ref{sec:planetselection}) except that we take full account
of finite source effects \citep{ob04343}, which are much more 
important for the present sample of events because of their 
higher magnification.  For the last three events above the
cut in Table~\ref{tab:events}, $\theta_\e$ (and hence 
$\rho\equiv\theta_*/\theta_\e$) is not well
measured. 
For these, we follow the procedure of \citet{gaudi02}
and adopt $\rho = \theta_*/(\mu_{\rm typ} t_\e)$, where
$\theta_*$ is determined from the instrumental color-magnitude
diagram in the standard way \citep{yoo04}, $t_\e$ is the measured
Einstein timescale, and $\mu_{\rm typ}=4\,\rm mas\,yr^{-1}$ is the
typical source-lens proper motion toward the Galactic bulge.
Figure \ref{fig:familyportrait} is a ``portrait
album'' of the
resulting triangle diagrams (only one side shown to conserve
space) and
Figure \ref{fig:sens} shows the integrated sensitivity
of each event as a function of mass ratio $q$.  That is, it is the
integral of the sensitivity (in Figs.\ \ref{fig:ob07050},
\ref{fig:ob08279}, and \ref{fig:familyportrait}) 
over horizontal slices.  Hence, if the
sensitivity were truly a triangle, the curves in Figure \ref{fig:sens}
would be perfectly straight lines, with slope $2\eta$ (illustrated by
the bold black line segment) and $x$-intercepts at $q_\min$.
Most of the curves do have this behavior over the range
$-4\la \log q\la -2$, which is the main range of sensitivity
of this technique and also where the planets in Table \ref{tab:parameters}
are located.  Moreover, the inferred intercepts of the
straight-line portion of these curves do generally reach to lower
mass ratio $q_\min$ for higher $A_\max$ events, although with
considerable scatter.  However, while some events (like OGLE-2008-BLG-279)
are almost perfectly straight down to zero, others 
(like OGLE-2005-BLG-169 and OGLE-2006-BLG-109)
show a sharp flattening toward lower mass ratios.  There are
two reasons for this. Events like OGLE-2005-BLG-169  have
non-uniform coverage over peak, which makes detectability a
strong function of angle.  The contours in the triangle diagram
separate, so that while the 50\%-sensitivity contour is fairly
straight, there is still substantial sensitivity below
$q_\min$, which is defined by where the two 50\% contours meet,
creating a long tail of sensitivity below this threshold.
Events like OGLE-2006-BLG-109 have very small source size 
relative to impact parameter, $\rho/u_0\ll 1$, which enables
detection of small mass-ratio planets that are very close to
Einstein ring because the relatively large, but very weak, caustics
of these planets are then not ``washed out'' as they would be
for larger $\rho$ \citep{bennettrhie96}.  
Hence the entire ``triangle'' has a curved
appearance, although the contours are tightly packed together.

The black bold dashed curve in Figure \ref{fig:sens}
is the combined sensitivity, i.e., the sum of the sensitivities for
all 13 events (divided by 10, so it fits on the same plot),
which we call $G(q)$.

The curves in Figure \ref{fig:sens} allow us to compare the
observed log projected separation $s$, with the maximum detectable separation.
See Table \ref{tab:parameters}.  Because some planets suffer from
the $s\leftrightarrow s^{-1}$ degeneracy, we only show the absolute
value of $\log s$.
The cumulative distribution of the ratios of these quantities is
shown in Figure \ref{fig:doverdmax}.  The separations are consistent with
being uniform in $\log s$, with Kolmogorov-Smirnov probability of 20\%.
(Moreover, in general, the high-magnification events with 
the greatest values of $|\log s|$ -- and so the greatest
potential leverage for probing the distribution as a function
of $s$ -- are also the most severely affected by the
$s\leftrightarrow s^{-1}$ degeneracy.  This applies, for example,
to MOA-2007-BLG-400Lb.  Thus the dependence on $s$ will be much better
explored using ``planetary caustics'' in low-magnification
events, for which the $s\leftrightarrow s^{-1}$ degeneracy is easily
resolved. \citealt{gouldloeb92,gaudigould97})

{\subsection{Masses of Host Stars}
\label{sec:hostmass}}

Now, $f(q)$  may in principle be a function of 
the host mass $M$ (and perhaps other variables as well).
With only six detections,
we are obviously in no position to subdivide our sample.  
Nevertheless, it is important to assess what host mass range
we are actually probing.  There do exist mass estimates or
limits for all 5 hosts of the planets that have been detected,
and there are also mass estimates for 5 of the 8 lensing stars
in the sample for which no planet was detected, which are given
in Table \ref{tab:events} together with the method of estimation.
For 5 of the events, both the angular Einstein radius $\theta_\e$
and the ``microlens parallax'' $\pi_\e$ are measured, which together permits
a mass measurement $M=\theta_\e/\kappa\pi_\e$, where
$\kappa\equiv 4G/(c^2{\rm AU})\sim 8.1\, {\rm mas}\,M_\odot^{-1}$
\citep{gould00b}.  The measurement for OGLE-2007-BLG-349 is
preliminary (Dong et al., in preparation) but the others
are secure.  For four of the events, there is a measurement
of $\theta_\e=\sqrt{\kappa M\pi_{\rm rel}}$, where $\pi_{\rm rel}$ is the
source-lens relative parallax, but not $\pi_\e$.  This measurement
of the product of $M$ and $\pi_\rel$ combined with the lens-source
relative proper motion
$\mu=\theta_\e/t_\e$ and a Galactic model (GM) permit a Bayesian
estimate of $M$.  Finally there are two events for which adaptive
optics (AO) observations provide information on the host.
For OGLE-2006-BLG-109, AO resolution of the host confirms the microlensing
mass determination from $\theta_\e$ and $\pi_\e$.  For MOA-2008-BLG-310,
AO observations detect excess light (not due to the source) but it
is not known whether this excess is due to the lens or another star.  So
only an upper limit on the mass is obtained.

Seven of these 10 measurements are in the range of middle M to 
middle K stars, while one is a brown dwarf and another is likely
to be a late M dwarf.  They cover a fairly broad range
approximately centered on $0.5\,M_\odot$, with a tail toward
lower mass. Of course,
there are also 3 lenses in the $A_\max>200$ sample for which there
is no mass measurement or estimate.  These have timescales $t_\e$ of
10, 26, and 59 days, which is much more typical of microlensing
events than the sample having mass measurements
 (with 3 events having $t_\e > 100\,$days).
If these are otherwise typical events, then the lenses probably lie mostly
in the Galactic bulge \citep{kiraga94}, in which case their mean mass
is roughly $0.4\,M_\odot$ \citep{gould00a}.  
Given that this is a minority of the
sample, that the information about this minority is far less secure, and
that the difference from the sample with harder information is not
very large, we adopt 
\begin{equation}
M\sim 0.5\,M_\odot
\label{eqn:midmass}
\end{equation} 
for the typical mass of the
sample.  However, we note that the implications discussed in 
Section~\ref{sec:discuss} would not be greatly affected if we had
adopted $M\sim 0.4\,M_\odot$.

{\subsection{Likelihood Analysis}
\label{sec:likelihood}}

To evaluate the mass-ratio distribution function $f(q) = Aq^n$,
we maximize the likelihood:
\begin{equation}
L = -N_{\rm exp} +\sum_{i=1}^{N_{\rm obs}}\ln G(q_i) f(q_i);\qquad
N_{\rm exp} = \int_{q_-}^{q_+} d q G(q)f(q)
\label{eqn:likelihood}
\end{equation}
and find
\begin{equation}
f(q) = {d N_\ev\over d\log q \,d\log s} = (0.36 \pm 0.15)\times 
\biggl({q\over 5\times 10^{-4}}\biggr)^{-0.60\pm 0.20}\,{\rm dex}^{-2}.
\label{eqn:fofq_eval}
\end{equation}
However, as we now argue, while the normalization of 
Equation (\ref{eqn:fofq_eval}) is robust, the slope is not.

In Section \ref{sec:planetselection}, we summarized why there 
must be some boundaries $q_\pm$ beyond which the experiment
is seriously degraded, but argued that there is no ``impartial
algorithm'' for deciding exactly where those boundaries should be.
We find that if we vary $q_+$ between $-2.3$ and $-2.$, and
we vary $q_-$ between $-5$  and $-4.5$, that the normalization
in Equation (\ref{eqn:fofq_eval}) varies by only $\pm 10\%$, which
is much smaller than the statistical errors.  However, the power-law index
varies between $-0.6$ and $-0.2$.  If we had a much larger sample
of planets, then we could set the boundaries at various places
within the range of our detections, thereby simultaneously
reducing both the number of detections and $N_{\rm exp}$ in
Equation (\ref{eqn:likelihood}).  For an infinite sample, such a 
procedure should lead to no variation in either slope or normalization.
For a finite sample, the variation would provide an estimate 
of the error in these quantities due to
the uncertainty in knowledge of these boundaries.  However, when
we apply this procedure to our small sample, we find quite wild
variations, implying that we cannot derive a reliable slope from these
data.

In Section \ref{sec:planetselection} we mentioned that the
threshold value $\Delta\chi^2>500$ also had some intrinsic
uncertainty.  However, in this case the effect is very small.
For example, if we decrease the threshold to $\Delta\chi^2>350$,
then the normalization in Equation (\ref{eqn:fofq_eval}) decreases by
only 7\%, much less than the Poisson error. Given that the normalization
in Equation (\ref{eqn:fofq_eval}) is robust but the slope is not, we
give our final result as:
\begin{equation}
{d N_\ev\over d\log q \,d\log s} = (0.36 \pm 0.15)
\,{\rm dex}^{-2}
\qquad  {\rm at}\ \ q \sim  5\times 10^{-4}
\label{eqn:fofq_eval2}
\end{equation}

Figure \ref{fig:nofq} summarizes the principal inputs to the modeling.
The top panel shows the cumulative
distribution of the detections.  The bottom panel shows the
sensitivity of the survey as function of planet mass, both
the analytic approximation derived in Section \ref{sec:estimate}
and the numerical determination derived in Section \ref{sec:fullsample}.
These hardly differ.

The density given in Equation (\ref{eqn:fofq_eval2}), 
can be obtained by a very simple
argument. The ``triangle'' for each event has sensitivity to
(1/2 base $\times$ height) $= \eta [\log (0.01\,A_\max/\xi)]^2
\sim 1.7[1 + 0.42\log(A_\max/400)]^2\,{\rm dex}^2$ of planet parameter
space.  We observed $N=13$ events and found six planets,
so $6/(13\times 1.7)\sim 0.27\,{\rm dex}^{-2}$, i.e. correct within
the statistical error.

{\section{Discussion}
\label{sec:discuss}}

We have presented the first measurement of the absolute frequency
of planets beyond the snow line over the mass-ratio range
$-4.5 \la \log q \la -2$.  The resulting planet frequency,
Equation (\ref{eqn:fofq_eval2}), can be understood
directly from the data and the ``triangle'' sensitivity diagrams.
The result
is applicable to a range of host masses centered near $M\sim 0.5\,M_\odot$.
The distribution is consistent
with being flat in log projected separation $s$.

{\subsection{Comparison with Previous Microlensing Results}
\label{sec:ulenscomp}}

\citet{ob05169} had earlier concluded that ``cool Neptunes are common''
based on one of the planets analyzed here (OGLE-2005-BLG-169Lb) and
another planet with similar mass ratio, OGLE-2005-BLG-390Lb
\citep{ob05390}, which had been detected through another channel:
followup observations of low magnification events.  OGLE-2005-BLG-390Lb
was actually recognized as a possible planetary event during the
planetary deviation, but detailed review of these communications
and their impact on the observing schedule shows that this ``feedback''
was not critical to robust detection.  Moreover,
at that time there had been no other
detections through this channel, so the \citet{ob05390} detection
could also be treated as a ``controlled experiment''. These could
then be combined to obtain
an absolute rate for ``cool Neptunes'', albeit with large errors.
However, with only two detections, \citet{ob05169} were not able to specify
the mass range of ``cool Neptunes'' and hence were not able
to express their results in units of dex$^{-2}$ as we have done here.
If we nevertheless, somewhat arbitrarily, say that the 
\citet{ob05390} and \citet{ob05169}
result applies to 1 dex in $\log q$, centered at the $q=8\times 10^{-5}$
of their two detections, and adopt their 0.4 dex interval in $\log s$,
their estimated density of cool Neptunes can be translated to
$0.95^{+0.77}_{-0.55}\,{\rm dex}^{-2}$ ($1\,\sigma$).  To make a fair
comparison with Equation (\ref{eqn:fofq_eval2}), it is necessary
to adopt some slope for the mass function in order to account for
the factor $\sim 6$ difference in mass.  If we adopt the
\citet{ob07368} slope of $n=-0.68$ from microlensing, 
then our prediction for this mass range would be $1.25\, {\rm dex}^{-2}$.
If we adopt the \citet{cumming08} slope of $n=-0.31$ from RV,
it would be $0.64\, {\rm dex}^{-2}$.  Either way, these are consistent.

\citet{ob07368} analyzed all 10 published microlensing planets,
including the 5 that we analyze here.
They approximated the sensitivity functions
of this heterogeneous sample by a single power law 
($\propto q^m$, $m=0.6\pm 0.1$)
and derived a power-law mass-ratio distribution
$dN_\pl/d\log q \sim q^n$, $n=-0.68\pm 0.2$.  
Since we are unable to derive a slope from our analysis, and they
do not derive a normalization, there can be no direct comparison
of results.

{\subsection{Comparison with Radial Velocity Results}
\label{sec:rvcomp}}

Based on an analysis of RV planets, \citet{cumming08} derive
a normalization of $0.035\,{\rm dex}^{-2}$,  a factor 
$10\pm 4$ smaller than
the one found here.  A factor 
$(5\times 10^{-4}/1.66\times 10^{-3})^{-0.31} = 1.5$ of this difference
is due to the fact that they normalize at higher mass ratio.  
The remaining factor $7\pm 3$ difference is most likely due to the
different star-planet separations probed by current microlensing
and RV experiments.  The \citet{cumming08} study targets stars
with periods of 2--2000 days, corresponding to a mean semi-major-axis
of $a=0.31\,$AU.
Microlensing probes a factor $\sim 3$ beyond the snow line
(Fig.\ 8 from \citealt{ob07368})\footnote{
Just as RV measurements respond to a projected stellar velocities,
and so measure $m\sin i$ of the planet which is always less than
or equal to the planet mass $m$, so
microlensing observations measure the projected separation $s$,
which for circular orbits is related to the semi-major axis by
$R_\e s = a\sin\gamma$ where $\gamma$ is the angle between the
star-planet axis and the line of sight. The statistical distribution
$\sin\gamma$ is exactly the same as for $\sin i$ in RV.  Hence,
except for rare cases when the orbit is constrained by higher-order
effects \citep{ob05071b,bennett10}, $a$ must be statistically estimated
from $s$ (and $R_\e$), which is what is done in Fig.\ 8 of \citet{ob07368}.
}.  
To make contact between
microlensing observations of primarily lower mass stars with RV observations
of typically solar type stars, we should consider planets in similar
physical conditions, which we choose to normalize by the snow line.
That is, we should compare to G-star planets at 3 ``snow-line radii'', i.e.,
$a\sim 8$ AU.
Hence, the inferred slope between the RV and microlensing measurements is 
$d\ln N/d\ln a = \log (7\pm 3)/\log(8/0.31) = 0.56\pm 0.16$, which is 
consistent with the slope of $d\ln N/d\ln a = 0.39\pm 0.15$ derived 
by \citet{cumming08} for RV stars {\it within} their period range.
Thus, simple extrapolation of the RV density profile derived from
planets thought to have migrated large distances, adequately predicts 
the microlensing results based on planets beyond the snow line that are 
believed to have migrated much less.  See Figure \ref{fig:semidist}.
Figure \ref{fig:qdist} compares microlensing and RV detections
as a function of mass ratio $q$.

{\subsection{Prospects for Sensitivity to Very Low Mass Planets}
\label{sec:verylowmass}}

Equation (\ref{eqn:gofq}) and Figure \ref{fig:sens} show a
break in sensitivity at $q_\thr \simeq 10^{-4}$.  For
a power-law mass-ratio distribution, the ratio of planets
expected above and below this threshold is
\begin{equation}
{N_\pl(q>q_\thr)\over N_\pl(q<q_\thr)} = {n+1\over n}
\biggl[z^n\ln(z) + \biggl({n-1\over n}\biggr)(z^n - 1)\biggr],
\label{eqn:ratio}
\end{equation}
where $z\equiv 0.01/q_\thr$.  This ratio is an extremely strong
function of the adopted slope of the mass function, $n$.
For $n=-0.68$ \citep{ob07368}, it is 1.0, whereas for
$n=-0.31$ \citep{cumming08}, it is 4.7.
The fraction of planets
within the lower domain that lies below $q$ is simply $q^{n+1}$.
Hence, while several authors have shown
that individual planets at or near Earth mass ratio are detectable
in high-magnification events \citep{abe04,ob04343,ob08279,ob07050}, 
the actual rate of detection will be strongly influenced by the
actual value of $n$.  As discussed in Section \ref{sec:planetselection},
probing to lower masses will require technical advances to robustly
identify low mass planets in the presence of higher mass planets.  
But it will also require increasing the number of events that
are monitored.

There is some potential to do this.
First, as shown in Section \ref{sec:numerical},
only about half of events that are announced by search teams
are intensively monitored.  Hence, there is room to double
the rate by more aggressive monitoring.  This will be aided
by inauguration of OGLE-IV, which will have much higher
time sampling and so will permit more accurate prediction
of high-mag events.  Second, 
it is possible that the more intensive OGLE-IV survey will
increase the underlying sample of high-mag events.
Finally, systematic 
analysis of high-mag events could bring down the effective $\Delta\chi^2$ 
threshold from 500 to, say, 200, which would decrease $\xi$ (and
so $q_\thr$) by a factor $(500/200)^{2/3}\sim 1.8$.  
This would bring only
a modest (logarithmic) increase in the sensitivity in the range $q>q_\thr$ 
but would aid linearly for $q<q_\thr$.

{\subsection{Constraints on Migration Scenarios}
\label{sec:migration}}

We showed in Section \ref{sec:rvcomp} that the planet density
derived here, $d N_\pl/d\log q d\log s = 0.36\,{\rm dex}^{-2}$ 
is consistent with the density derived from RV studies, if the latter
are extrapolated to $\sim 25$ times the semi-major axis where the
measurement is made.  Regardless of the details of this comparison,
the fact that the density of planets beyond the snow line is 7 times
higher than that at 0.3 AU, indicates that {\it most giant planets
do not migrate very far}.  Moreover, the fact that the slope found
in RV studies at small $s$ adequately predicts the density at large
$s$, would seem to imply that whatever is governing the amount
of migration is a {\it continuous} parameter.  That is, it is not
the case that there are two classes of planetary systems: those
with migration and those without.  Rather, all systems have migration,
but by a continuously varying amount.  This picture would be in 
accord with the evolving view of the Solar System, that even though
the giant planets are in the general area of their birth ``beyond the
snow line'', they have migrated to a modest degree.

{\subsection{Comparison to Solar System}
\label{sec:ss}}

Another interesting point of comparison is to the planet density in
the Solar System, where there are 4 planets in the mass-separation
regime that microlensing currently explores, 
Jupiter, Saturn, Uranus, and Neptune.  How common are 
``Solar-System analogs'', i.e., systems 
with several giant planets out beyond the snow line?

To address this question, we ask what the result of our study
would have been if every microlensed star possessed a ``scaled version''
of our own Solar System in the following sense: 4 planets with the 
same planet-to-star mass ratios and same ratios of semi-major
axes as the outer Solar-System planets, but with the overall
scale determined by the ``snow line''.  While there is 
observational evidence from the asteroid belt that the Solar-System snow
line is near $R_{{\rm snow},\odot}=2.7\,$AU \citep{morbidelli00},
there is considerable uncertainty on how this 
scales with stellar mass \citep{sasselov00,kennedy08}.
We therefore parametrize this
relation by
\begin{equation}
R_{\rm snow}(M) = R_{{\rm snow},\odot}\biggl({M\over M_\odot}\biggr)^\nu,
\label{eqn:snow}
\end{equation}
and consider a range $0.5< \nu < 2$, adopting $\nu=1$ for our fiducial
value.  We consider that the typical Einstein radius is
$R_\e = 3.5\,{\rm AU}\,(M/M_\odot)^{1/2}$ and the typical lensed star is
$M=0.5\,M_\odot$.  Then scaling down the semi-major axis of a Jupiter analog
so $a_{\rm Jup-analog}/R_{\rm snow} = a_{\rm Jup}/R_{{\rm snow},\odot}$
implies
\begin{equation}
{a_{\rm Jup-analog}\over R_\e} = {a_{\rm Jup}\over 3.5\,{\rm AU}}
\biggl({M\over M_\odot}\biggr)^{\nu-0.5},
\label{eqn:snow2}
\end{equation}
and similarly for the other 3 planets.  We then imagine
that these systems are viewed at random orientations, with
the individual planets in random phases.  From a Monte Carlo
simulation, we find that we would then have expected (for our fiducial
$\nu=1$) to have detected
18.2 planets 
(11.4 ``Jupiters'', 6.4 ``Saturns'', 0.3 ``Uranuses'', 0.2 ``Neptunes'') 
including 6.1 systems with 2 or more planet detections.  See Figure
\ref{fig:ss}.
For $1<\nu<2$, the planet totals and multi-planet detections barely
change.  However they fall somewhat
for smaller $\nu$, reaching 15.3 planets and 3.8 two-planet systems at
$\nu=0.5$.

These results compare
to 6 planets including 1 two-planet system in the actual sample.

Hence, our Solar System appears to be 3 times richer in planets
than other stars along the line of sight toward the Galactic bulge.
The single detection of a multi-planet system \citep{ob06109}
allows the first estimate of the frequency of stars with
``solar-like systems'', defined as having multiple giants in the
snow zone: 1/6, albeit with large errors.

\acknowledgments

\acknowledgements{
Work by AG was supported in part by NSF grant AST 0757888 and in part
by NASA grant 1277721 issued by JPL/Caltech. Work by SD was performed
under contract with the California Institute of Technology (Caltech)
funded by NASA through the Sagan Fellowship Program. This work was
supported in part by an allocation of computing time from the Ohio
Supercomputer Center. The OGLE project is partially supported by the
Polish MNiSW grant N20303032/4275 to AU.The MOA collaboration was
supported by the Marsden Fund of New Zealand. The MOA collaboration
and a part of authors are supported by the Grant-in-Aid for Scientific
Research, JSPS Research fellowships and the Global COE Program "Quest
for Fundamental Principles in the Universe" from JSPS and MEXT of
Japan. TS acknowledges support from grants JSPS18749004, JSPS20740104,
and MEXT19015005. The RoboNet project acknowledges support from PPARC
(Particle Physics and Astronomy Research Council) and STFC (Science
and Technology Facilities Council). CH was supported by Creative
Research Initiative Program (2009-0081561) of National Research
Foundation of Korea. AGY's activity is supported by a Marie Curie IRG
grant from the EU, and by the Minerva Foundation, Benoziyo Center for
Astrophysics, a research grant from Peter and Patricia Gruber Awards,
and the William Z. and Eda Bess Novick New Scientists Fund at the
Weizmann Institute. DPB was supported by grants AST-0708890 from the
NSF and NNX07AL71G from NASA. Work by SK was supported at the Technion
by the Kitzman Fellowship and by a grant from the Israel-Niedersachsen
collaboration program. Mt Canopus observatory is financially supported
by Dr David Warren.}

\vfil\eject

%\documentclass[12pt,preprint]{aastex}
%\usepackage{lscape}
%\usepackage{rotating}
%\begin{document}

\begin{table}
\caption{\label{tab:events} \sc Monitored Events with Magnification $A>100$}
\vskip 1em
\begin{tabular}{@{\extracolsep{0pt}}lrcrll}
\hline
\hline
Name & $A_{\rm max}$ & $t_0 $(HJD) & $t_\e$ & $M/M_\odot$ & Method \\ \hline
\hline
OGLE-2007-BLG-224 &2424 & 4233.7 &   7 & $0.056\pm 0.004$ & $M=\theta_\e/\kappa\pi_\e$ \\ \hline
OGLE-2008-BLG-279 &1600 & 4617.3 & 101 & $0.64\pm0.10$ & $M=\theta_\e/\kappa\pi_\e$ \\ \hline
OGLE-2005-BLG-169 & 800 & 3491.9 &  43 & $0.49^{+0.23}_{-0.29}$ & GM$\oplus\theta_\e\oplus t_\e$ \\ \hline
 MOA-2007-BLG-400 & 628 & 4354.6 &  14 & $0.30^{+0.19}_{-0.12}$ & GM$\oplus\theta_\e\oplus t_\e$ \\ \hline
OGLE-2007-BLG-349 & 525 & 4348.6 & 121 & $\sim 0.4$ & $M=\theta_\e/\kappa\pi_\e$ \\ \hline
OGLE-2007-BLG-050 & 432 & 4222.0 &  68 & $0.50\pm 0.14$ & $M=\theta_\e/\kappa\pi_\e$ \\ \hline
 MOA-2008-BLG-310 & 400 & 4656.4 &  11 & $\leq 0.67\pm 0.14$ & AO \\ \hline
OGLE-2006-BLG-109 & 289 & 3831.0 & 127 & $0.51^{+0.05}_{-0.04}$ & $M=\theta_\e/\kappa\pi_\e$,AO \\ \hline
OGLE-2005-BLG-188 & 283 & 3500.5 &  14 & $0.16^{+0.21}_{-0.08}$ & GM$\oplus\theta_\e\oplus t_\e$ \\ \hline
 MOA-2008-BLG-311 & 279 & 4655.4 &  18 & $0.20^{+0.26}_{-0.09}$ & GM$\oplus\theta_\e\oplus t_\e$ \\ \hline
 MOA-2008-BLG-105 & 267 & 4565.8 &  10 &  &  \\ \hline
OGLE-2006-BLG-245 & 217 & 3885.1 &  59 &  &  \\ \hline
OGLE-2006-BLG-265 & 211 & 3893.2 &  26 &  &  \\ \hline \hline
OGLE-2007-BLG-423 & 157 & 4320.3 &  29 &  &  \\ \hline 
OGLE-2005-BLG-417 & 108 & 3568.1 &  23 &  &  \\ \hline

\end{tabular}
\end{table}
%\end{document}

%OGLE-2007-BLG-514 & 170 & 4386.5 & 0.9 \\ \hline
%OGLE-2008-BLG-270 & 500 & 4603.5 & 1.0 \\ \hline
%OGLE-2008-BLG-290 &  97 & 4632.6 & 1.0 \\ \hline
%MOA-2008-BLG-383  & 240 & 4688.5 & 0.4 \\ \hline

%\documentclass[12pt,preprint]{aastex}
%\usepackage{lscape}
%\usepackage{rotating}
%\begin{document}

\begin{table}
\caption{\label{tab:parameters} \sc Planets in Densely Monitored High-mag Events}
\vskip 1em
\begin{tabular}{@{\extracolsep{0pt}}lccccc}
\hline
\hline
Name               & {$\log u_0 $} & {$\log \rho $} & {$\log q $} & {$|\log s |$} & {$|\log s_{\rm max}|$} \\ \hline
\hline
%OGLE-2005-BLG-071Lb & -1.6         & -3.4          & -2.1       & 0.11         & 0.41 \\ \hline
OGLE-2005-BLG-169Lb & -2.9         & -3.4          & -4.1       & 0.009        & 0.19 \\ \hline
OGLE-2006-BLG-109Lb & -2.5         & -3.5          & -2.9       & 0.20         & 0.39 \\ \hline
OGLE-2006-BLG-109Lc & -2.5         & -3.5          & -3.3       & 0.017         & 0.27 \\ \hline
OGLE-2007-BLG-349Lb & -2.7         & -3.3          & -3.5       & 0.099         & 0.48 \\ \hline
MOA-2007-BLG-400Lb & -3.6         & -2.5          & -2.6       & 0.47         & 0.55 \\ \hline
MOA-2008-BLG-310Lb & -2.5         & -2.3          & -3.5       & 0.035        & 0.14 \\ \hline
\end{tabular}
\end{table}
%\end{document}

\begin{figure}
\plotone{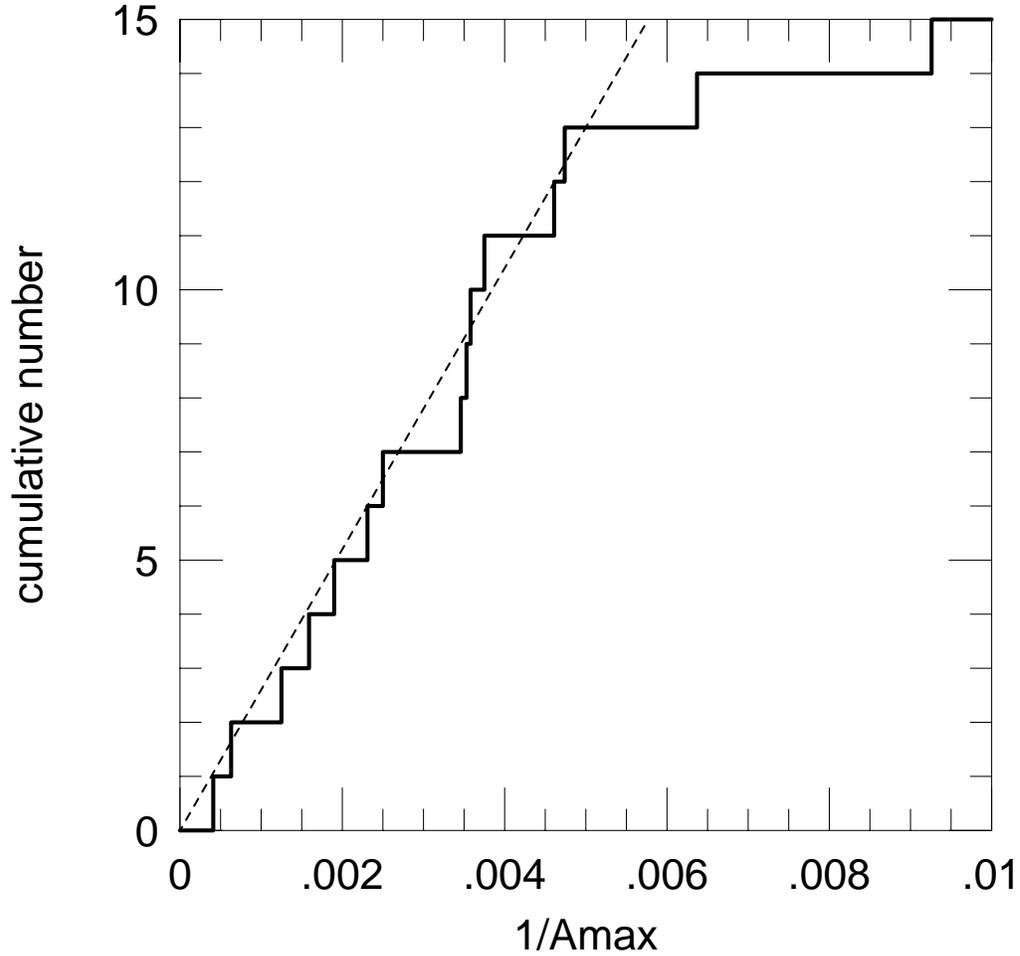}
\caption{\label{fig:amax}
Cumulative distribution of inverse maximum magnification
$A_{\rm max}^{-1}$ for high-mag
events observed by $\mu$FUN during the years 2005-2008.  
 The distribution is uniform in $A_{\rm max}^{-1}$ 
for high-mag events $A_{\rm max}>200$, with a slope of 
$dN_\ev/d A_\max^{-1}=2600$ ({\it dashed line}).
A ``controlled experiment'' therefore requires a selection
criterion $A_\max>200$.  See Section \ref{sec:eventselection}.
}
\end{figure}

\begin{figure}
\plotone{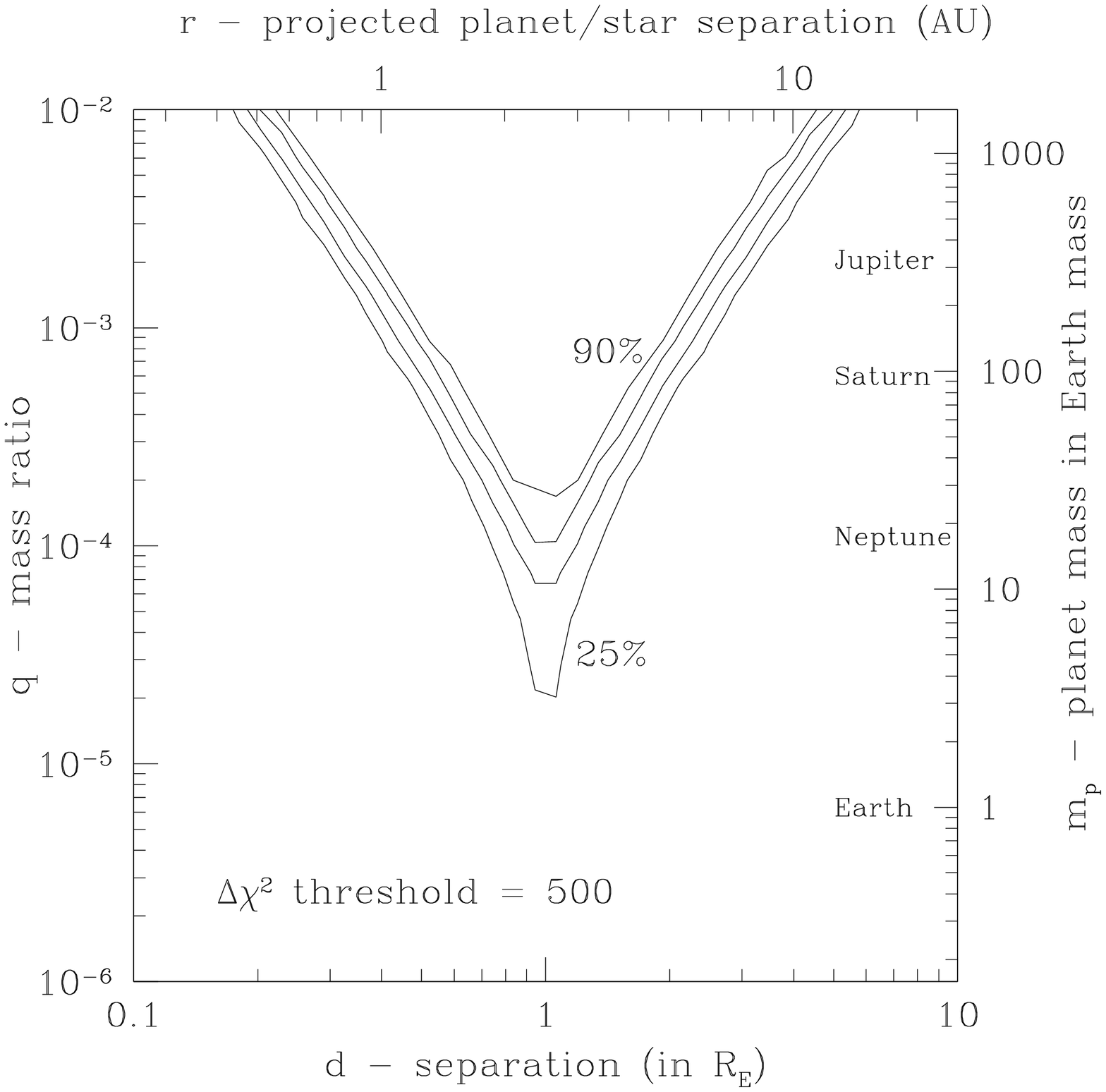}
\caption{\label{fig:ob07050}
Triangle diagram for OGLE-2007-BLG-050 (adapted from
Fig.\ 9 of \citealt{ob07050})
showing sensitivity to planets of planet/star mass ratio $q$ and
planet-star projected separation $d$ (alias $s$ in the current paper).
Planets within the contours would be detectable at $\Delta\chi^2>500$
for 25\%, 50\%, 75\% and 90\% of source-lens trajectories.  Since
these trajectories are random, the contours reflect the probability
of detecting the planet at a given $s$ and $q$.  The slope of
the contours is $\eta = 0.32$. 
}
\end{figure}

\begin{figure}
\plotone{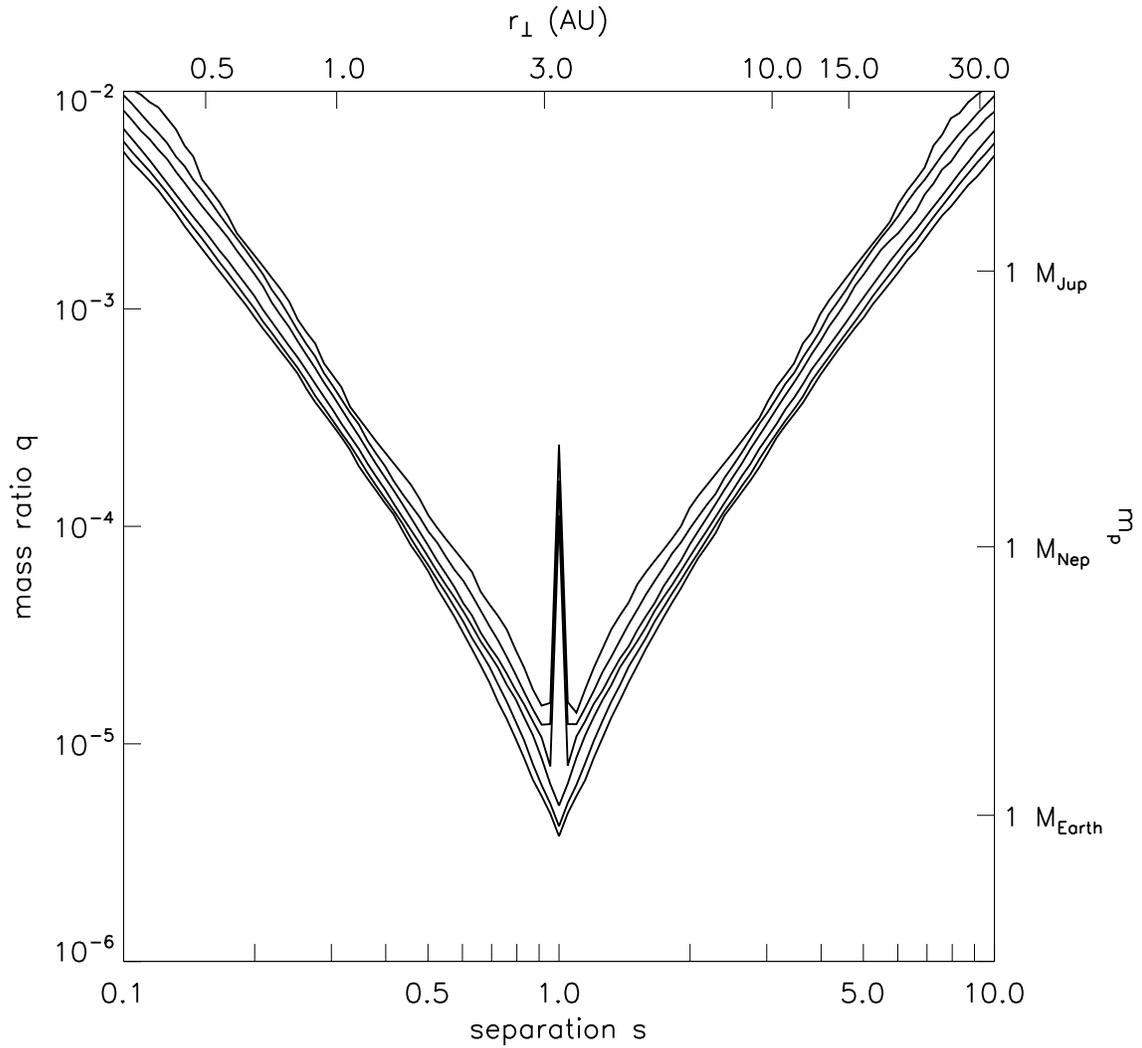}
\caption{\label{fig:ob08279}
Triangle diagram of planet sensitivities for OGLE-2008-BLG-279 (adapted from
Fig.\ 7 of \citealt{ob08279}).
Similar to Fig.\ \ref{fig:ob08279} except with
contours of $\Delta\chi^2>500$ detectability with
10\%, 25\%, 50\%, 75\% 90\% and 99\% probability.
Contour slope is $\eta = 0.35$.
}
\end{figure}

\begin{figure}
\plotone{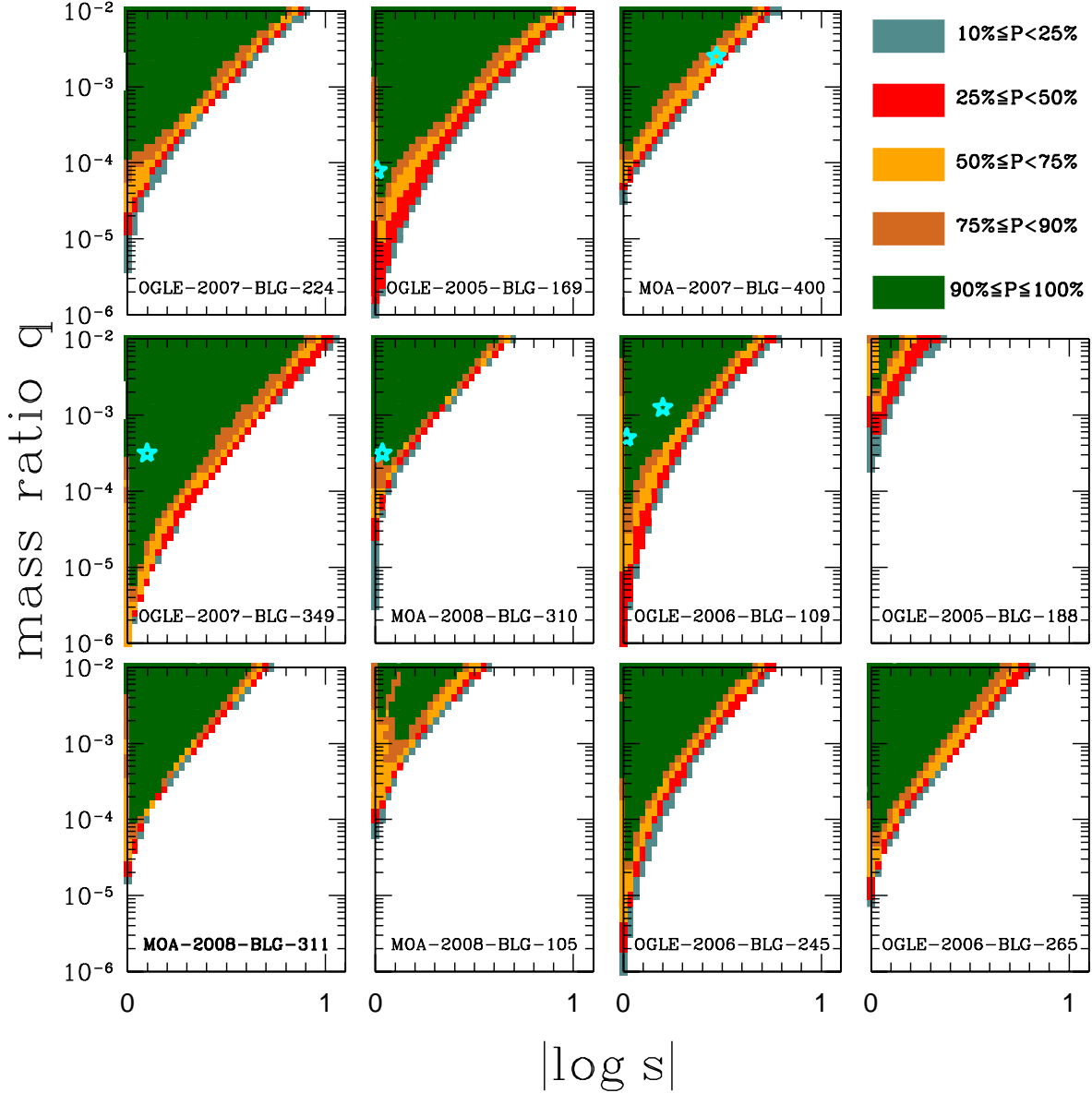}
\caption{\label{fig:familyportrait}
New ``triangle diagram'' sensitivities for all events in our sample except
the two shown in Figures \ref{fig:ob07050} and \ref{fig:ob08279}.
Sensitivity is the fraction of all trajectory angles $\alpha$ that
a planet would have been detected at $\Delta\chi^2>500$, if a planet
of mass ratio $q$ had been present at projected separation $s$ (in
units of the Einstein radius).  Only half of the diagram
(which is almost perfectly symmetric) is shown here to conserve
space.  Positions $(\log q,|\log s|)$ of all detected planets
from Table~\ref{tab:parameters} are shown as gold stars.
}
\end{figure}

\begin{figure}
\plotone{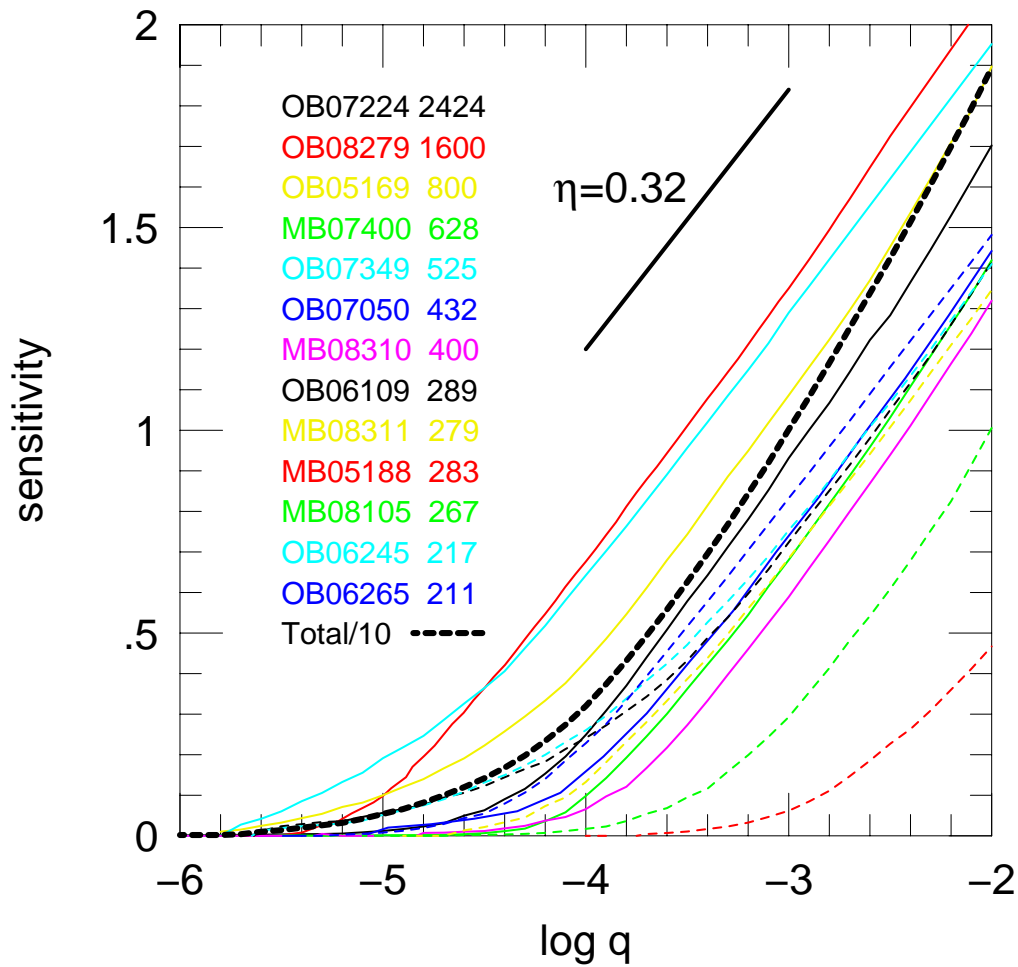}
\caption{\label{fig:sens}
Calculated sensitives for each of the 13 microlensing events.  These
are integrals over horizontal cuts (at fixed $q$) in Figures
\ref{fig:ob07050}, \ref{fig:ob08279}, and \ref{fig:familyportrait}.
Events are shown in ``rainbow order'' according to magnification,
with the first 7 in solid lines and the last 6 in dashed lines.
The bold dashed black curve represents the total sensitivity of
the sample (divided by 10).
}
\end{figure}

\begin{figure}
\plotone{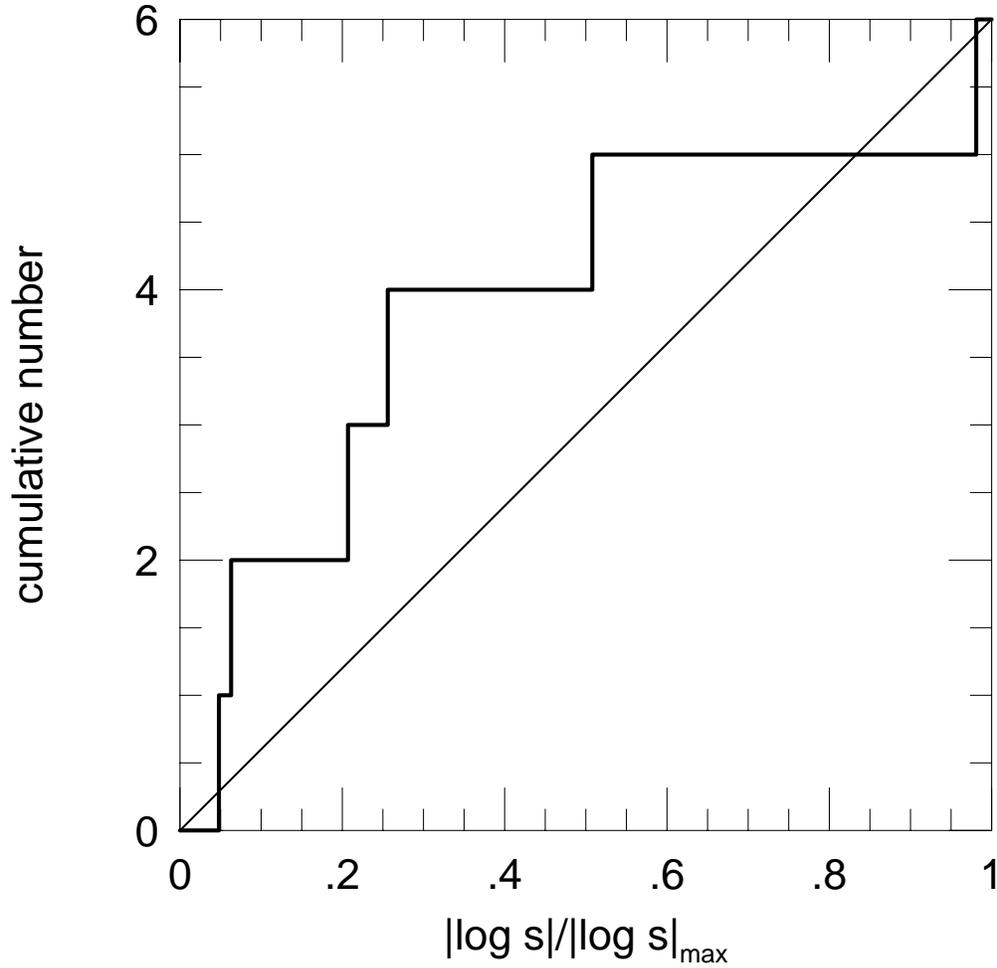}
\caption{\label{fig:doverdmax}
Cumulative distribution of the (absolute value of the log of the)
measured projected separation, $|\log s|$, in units of the
maximum value of this parameter,  derived in Section \ref{sec:fullsample}
The
observed distribution is consistent with planets being distributed
uniformly in $\log s$, with Kolmogorov-Smirnov probability of 20\%.
}
\end{figure}

\begin{figure}
\plotone{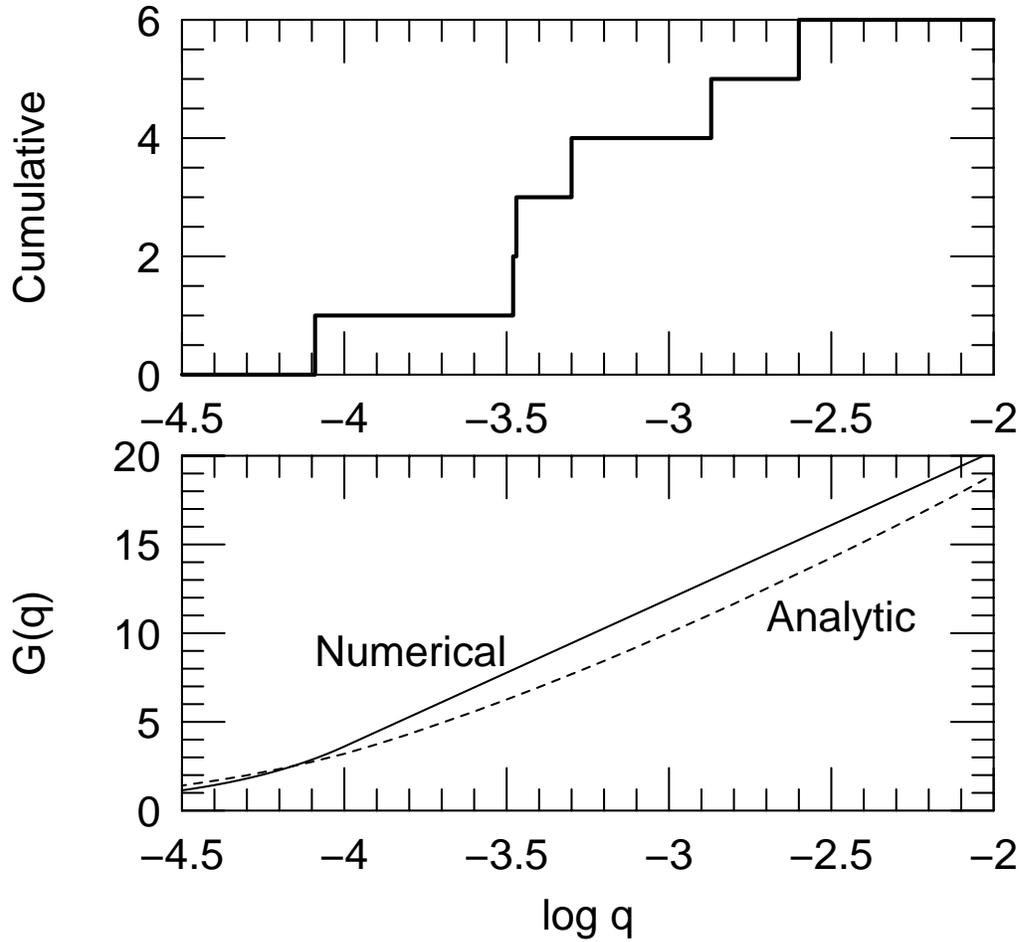}
\caption{\label{fig:nofq}
Principal inputs to the modeling.
Top panel: cumulative
distribution of planet detections in 4 years of intensive monitoring
of high-magnification events.
Bottom panel:
Sensitivity of the survey as a function of planet/star mass ratio
$q$, both
the analytic approximation derived in Section \ref{sec:estimate}
and the numerical determination derived in Section \ref{sec:fullsample}
and shown in Figure \ref{fig:sens}.
These hardly differ.
}
\end{figure}

\begin{figure}
\plotone{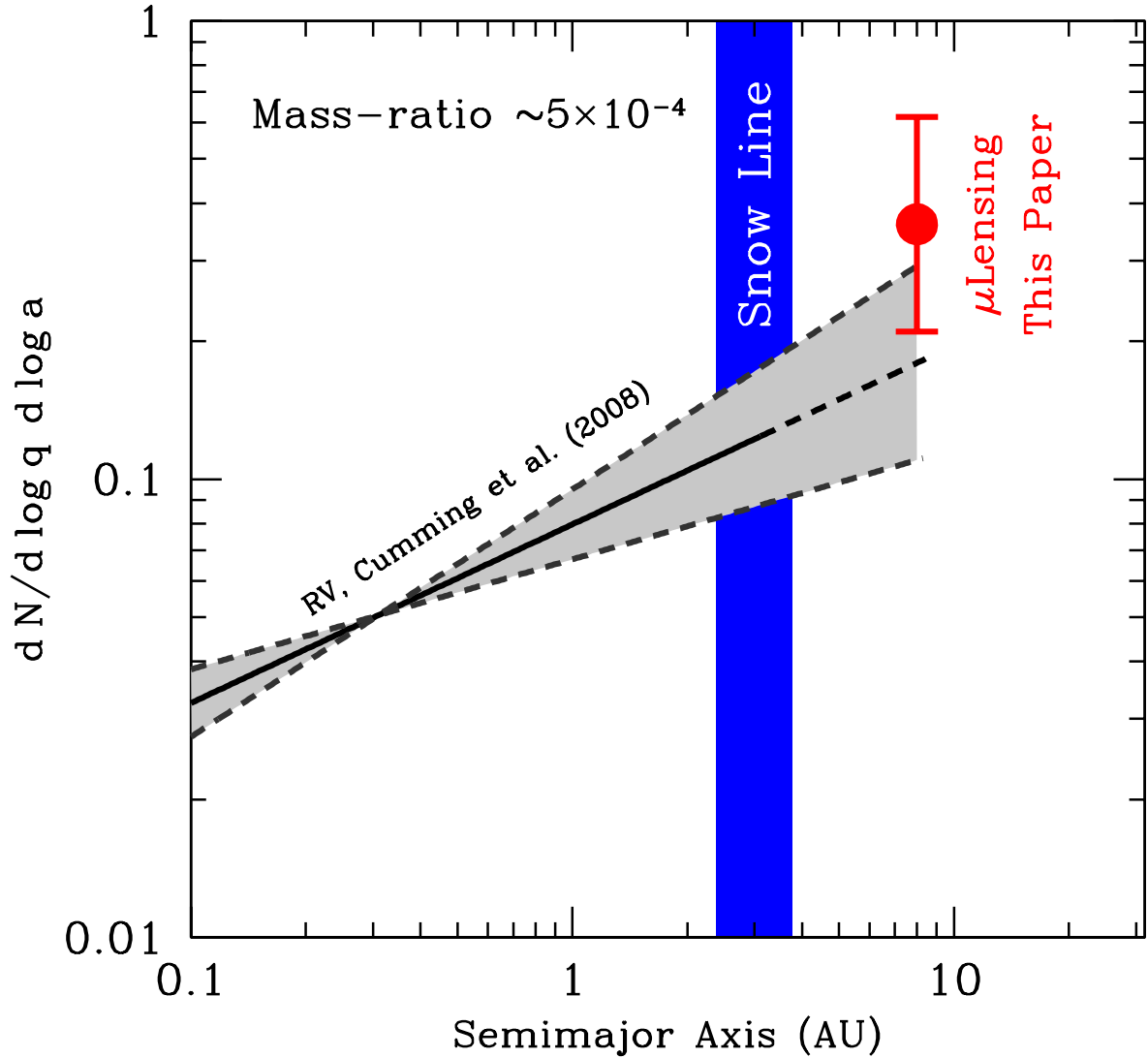}
\caption{\label{fig:semidist}
Planet frequencies determined from microlensing (this paper) and RV 
\citep{cumming08} at different semi-major axes.  The RV result is scaled
to mass ratio $q=5\times 10^{-4}$, using the RV-derived slope, $n=-0.31$.  
In order to take account of the different
host star masses ($M\sim 1\,M_\odot$ for RV, $M\sim 0.5\,M_\odot$
for microlensing) we have placed the microlensing point at
$8\,$AU, i.e., 3 times the solar-system ``snow line'' distance.
This is because microlensing planets are typically detected
at 3 times the distance of their own systems' snow line (which
is of course much closer than 8 AU).
}
\end{figure}

\begin{figure}
\plotone{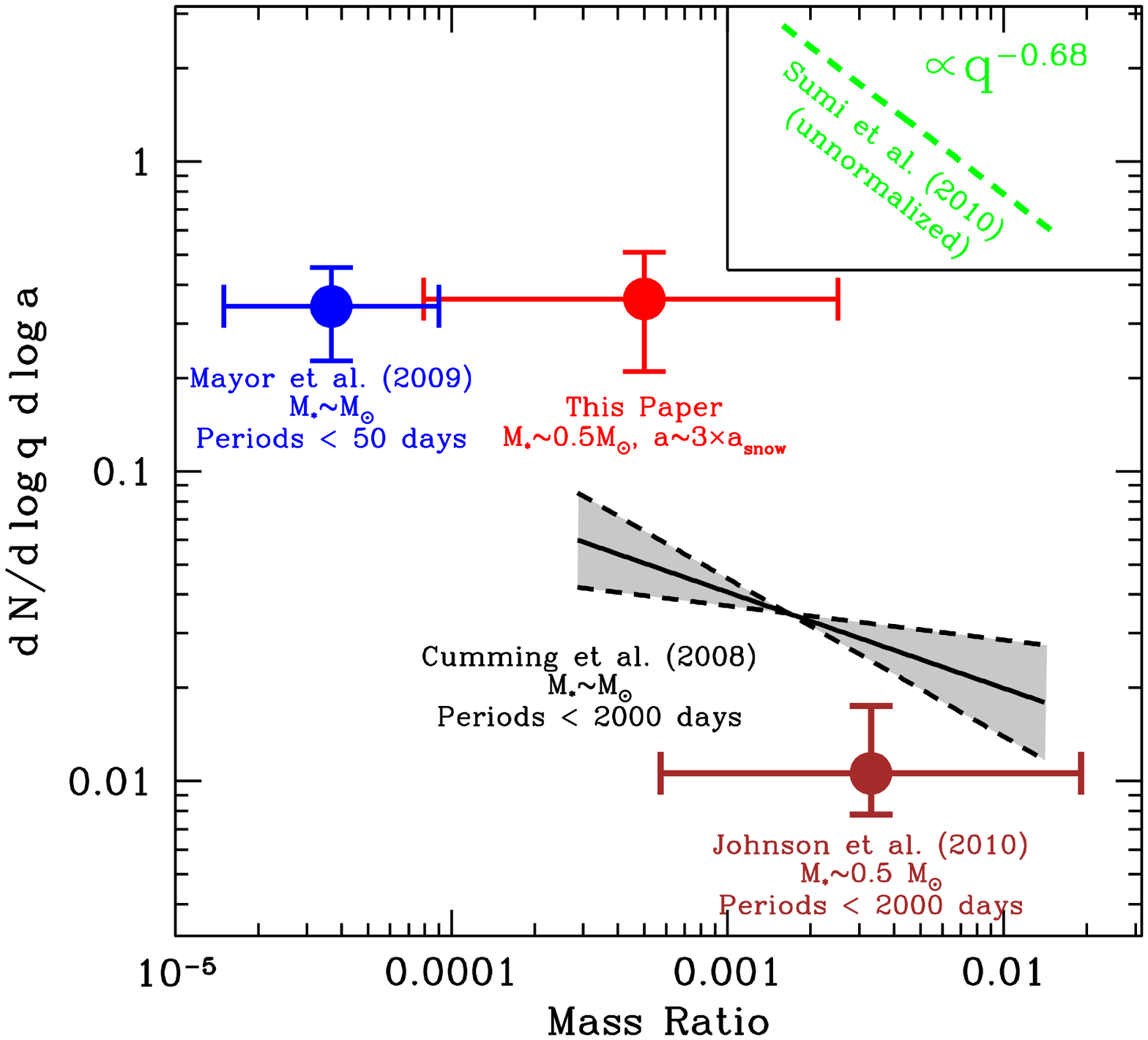}
\caption{\label{fig:qdist}
Planet frequencies as a function of mass ratio for microlensing
(this paper) and RV \citep{cumming08,mayor09,johnson10} detections.
As illustrated in Fig.\ \ref{fig:semidist}, the \citet{cumming08}
RV sample and the microlensing sample are consistent with each
other (despite different frequencies in this figure) because they are at 
different distances.  The \citet{cumming08} and
\cite{mayor09} RV samples are directly comparable on this diagram
because both are G stars, and they are consistent.
However, there is some tension between the \citet{johnson10} RV
measurement and the microlensing measurement, since both are
similar type stars.  This is because \citet{cumming08} and
\citet{johnson10} are ``inconsistent'' with each other, if
one assumes that the frequency of planets as a function of mass
ratio is independent of host mass, as we have done in arguing
for the consistency of microlensing with \citet{cumming08}.
Also shown is the (unnormalized) slope derived from microlensing
observations by \citep{ob07368}.
}
\end{figure}

\begin{figure}
\plotone{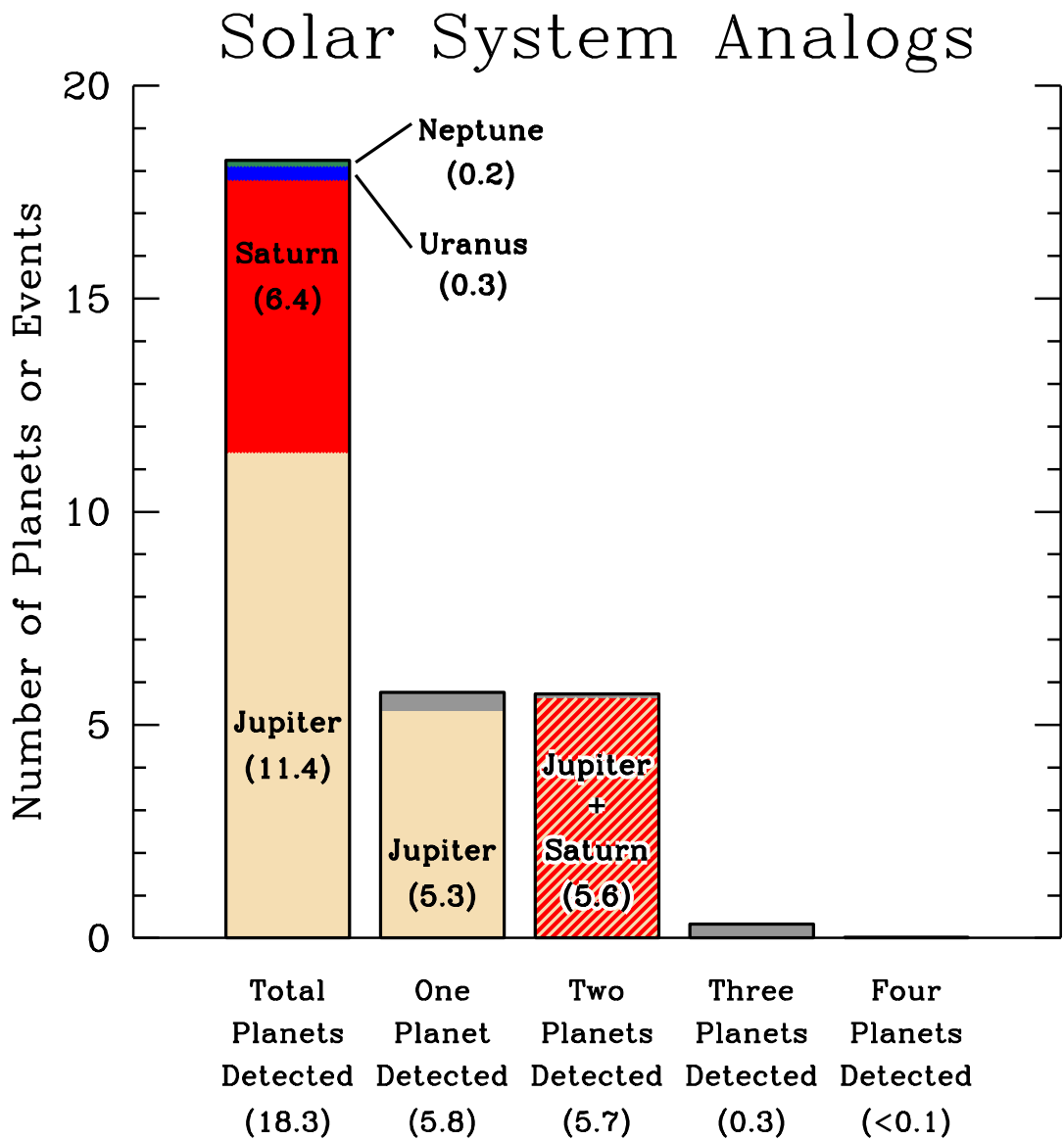}
\caption{\label{fig:ss}
Results of Monte Carlo simulation of the anticipated results
of our microlensing survey assuming that every star toward 
the Galactic bulge had a ``scaled solar system'',
as specified by Eqs.\ (\ref{eqn:snow}) and (\ref{eqn:snow2}).
A total of 18 planets would have been detected, including
6 two-planet systems, compared to the actual detections
of 6 planets including 1 two-planet system.  This seems to
indicate that the solar system is overdense in planets, especially
multiple planets.  Also shown are the frequency of various specific
combinations.
}
\end{figure}

\end{document}